\documentclass[a4paper,titlepage,12pt]{article}
\pdfoutput = 1

\usepackage[utf8]{inputenc}
\usepackage{color,graphicx}
\usepackage{epsfig}
\usepackage{amsmath}
\usepackage{verbatim}
\usepackage{amssymb}
\usepackage{mathabx}
\usepackage{stmaryrd}
\usepackage{empheq}
\usepackage{esint}
\usepackage{physics}
\usepackage{amsfonts}
\usepackage{cite}
\usepackage{array}
\usepackage{setspace}
\usepackage{braket}
\usepackage{float}
\usepackage{url}
\usepackage{mathtools}
\usepackage{tensor}
\usepackage{bbold}
\usepackage{slashed}
\usepackage{tikz}
\usepackage{graphicx}
\usepackage{epstopdf}
\usepackage[
  font=normalsize,
  labelfont=bf,
  textfont=normalfont,
  justification=justified,
  singlelinecheck=false
]{caption}
\usepackage{subcaption}
\usepackage{pgfplots}
\usepackage{mathrsfs}
\usepackage{fancybox}
\usepackage{eurosym}
\usepackage{tcolorbox}
\usepackage{tensor}
\allowdisplaybreaks
\usepackage{dsfont}
\usepackage{simpler-wick}
\usepackage{graphicx}
\usetikzlibrary{positioning,arrows.meta}

\usepackage[margin = 2.2cm]{geometry}
\usepackage[ragged]{footmisc}
\usepackage[bookmarks=true,bookmarksnumbered=false,
hyperindex=true,bookmarksopen=true,hyperfigures=true,
colorlinks=true,linkcolor=webblue,citecolor=webgreen,urlcolor=webblue,breaklinks]{hyperref}
\definecolor{webgreen}{rgb}{0, 0.5, 0}
\definecolor{webblue}{rgb}{0, 0, 0.5}
\definecolor{webred}{rgb}{0.5, 0, 0}
\definecolor{darkgreen}{rgb}{0,0.5,0}
\definecolor{darkblue}{RGB}{0,0,139}

\def\ben{\begin{equation}}
\def\een{\end{equation}}

   \let\d=\delta 
   
\let\l=\lambda     \let\r=v

\def\be{\begin{equation}}
\def\ee{\end{equation}}
\def\ba{\begin{array}}
\def\ea{\end{array}}

\def\dalemb#1#2{{\vbox{\hrule height .#2pt
       \hbox{\vrule width.#2pt height#1pt \kern#1pt
               \vrule width.#2pt}
       \hrule height.#2pt}}}

\newcommand{\bea}{\begin{eqnarray}}
\newcommand{\eea}{\end{eqnarray}}

\def\R{{{\mathbb{R}}}}

\let\tilde=\widetilde

\usepackage{xcolor}
\definecolor{greenUC}{RGB}{87,195,2}

\newcommand{\eps}{\varepsilon}

\renewcommand{\d}{\mathrm{d}}

\numberwithin{equation}{section}

\begin{document}

\thispagestyle{empty}
\begin{center}
    ~\vspace{5mm}

     {\LARGE \bf 
     Quantum gravity around ultracold black holes 
     from DSSYK 
        
   }
    
   \vspace{0.4in}
    

    

{ Francesca Mariani,\footnote{{\protect\path{francesca.mariani@ugent.be}}} Thomas G. Mertens,\footnote{{\protect\path{thomas.mertens@ugent.be}}} Jacopo Papalini,\footnote{{\protect\path{jacopo.papalini@ugent.be}}} Thomas Tappeiner,\footnote{{\protect\path{thomas.tappeiner@ugent.be}}}}
\\[0.3cm]
{\normalsize { \sl Department of Physics and Astronomy
\\[1.0mm]
Ghent University, Krijgslaan, 281-S9, 9000 Gent, Belgium}}\\[3mm]
\end{center}

 \vspace{0.2cm}
{\small  \noindent 
\begin{center} 
\textbf{Abstract}
\end{center} 
\noindent We propose a quantum description of the ultracold Reissner-Nordstr\"om de Sitter black hole via a specific flat space limit of the double-scaled SYK model. Fluctuations in this regime reduce to flat JT gravity, which is ill-behaved as a quantum system. We show that including the first subleading correction to the dilaton potential breaks the leading order Hagedorn degeneracy but does not resolve the structural problem. Finally, we leverage the gauged description of the sine dilaton gravity model to propose a novel quantization of the near-horizon dynamics of ultracold black holes. This leads to a new proposal for the partition function where states in the asymptotics of the spectrum are suppressed dynamically, and features of  a discretized spacetime arise. We obtain a non-perturbative low temperature behavior of the partition function, which is quite different than the cold and Nariai physics. Our result can be interpreted as a dynamical way to enforce cosmic censorship on the quantum system.

}
 \vspace{0.3cm}
\vfil
\begin{flushleft}
\today
\end{flushleft}


\pagebreak
\setcounter{page}{1}
\tableofcontents

\newpage
\section{Introduction}
Near-extremal black holes offer a powerful window into quantum gravity, as their low temperatures enhance quantum effects in the near-horizon region. In particular, they exhibit universal infrared quantum effects that are largely insensitive to the details of the ultraviolet completion of the underlying gravitational theory. As a consequence, these systems have become an important laboratory for investigating the interplay between black hole physics and quantum gravity \cite{Iliesiu:2020qvm,Heydeman:2020hhw,Rakic:2023vhv,Kapec:2023ruw}.

The universality of such systems is understood from the solvability of the spherically symmetric ($s$-wave) sector of the geometry that dominates the dynamics. This justifies approaching the problem via dimensional reduction over the transverse geometry, leading to an emergent two-dimensional dilaton gravity theory, capturing the low-energy near-horizon dynamics of the original higher-dimensional black hole. In this context, the study of Jackiw-Teitelboim (JT) dilaton gravity \cite{Jackiw:1984je, Teitelboim:1983ux}, has become a major research direction in recent years \cite{Mertens:2022irh}.
Specifically, it can be shown that in systems with two horizons, with a prominent example being the Reissner-Nordstr\"om black holes in four dimensions, the near-extremal, near-horizon limit leads to the emergence of an $\mathrm{AdS}_2$ throat governed by JT gravity with negative cosmological constant.
More precisely such systems develop an infinitely long throat where the spacetime factorizes into an $\mathrm{AdS}_2 \times S^2$ geometry, with the $\mathrm{AdS}_2$ factor corresponding to the spacetime solution of JT gravity, which can in turn be described by an emergent boundary reparametrization mode \cite{Jensen:2016pah, Maldacena:2016upp, Engelsoy:2016xyb}. 

The possibility of performing an exact quantization of the Schwarzian theory \cite{Stanford:2017thb,Mertens:2017mtv,Bagrets:2016cdf,Yang:2018gdb,Blommaert:2018oro,Iliesiu:2019xuh,Mertens:2022irh}, together with its exact duality with JT gravity, has provided an invaluable tool for extracting quantum information about the corresponding near-extremal four-dimensional black hole. In particular, the partition function of near-extremal Reissner-Nordström black holes was computed, revealing that the residual semiclassical entropy is removed by the IR quantum effects associated with the JT gravity degrees of freedom \cite{Iliesiu:2020qvm,Heydeman:2020hhw}. Moreover, the exact knowledge of boundary correlation functions in JT gravity has served as a fundamental tool in characterizing the evaporation process of near-extremal charged black holes, as well as in the computation of scattering amplitudes and absorption cross sections \cite{Brown:2024ajk,Lin:2025wof,Maulik:2025hax,Emparan:2025sao,Biggs:2025nzs,Emparan:2025qqf, Betzios:2025sct}.

There is, however, another powerful perspective on the Schwarzian theory that has attracted considerable attention in recent years. Namely, the Schwarzian action arises as the low-energy large $N$ effective description of the Sachdev-Ye-Kitaev (SYK) model \cite{Kitaevtalks,Maldacena:2016hyu,Kitaev:2017awl}. 
Despite the fact that the same Schwarzian theory emerges in these two very different settings, these two realizations have so far interacted only indirectly. JT gravity has played the role of an intermediate bridge, allowing the near-horizon dynamics of near-extremal Reissner-Nordström black holes and the soft mode of the SYK model to be described within a common framework. 

A natural question that therefore arises is whether the correspondence can be extended beyond the Schwarzian regime. More specifically, one may ask whether the full SYK model can serve as a microscopic model for higher-dimensional black holes. This possibility is particularly intriguing since the full SYK model is expected to capture certain features of de Sitter physics: in particular, the presence of a maximum entropy state and an entropy profile that subsequently decreases as a function of the energy, suggesting the presence of a finite-dimensional Hilbert space. 
Although the full SYK model is not exactly solvable for generic values of its parameters, a solvable regime was identified in the form of the double-scaled SYK model \cite{Berkooz:2018qkz,Berkooz:2018jqr,Lin:2022rbf}. In this limit, both $N$, the total number of Majorana fermions, and $p$, the number of interacting fermions appearing in the Hamiltonian, are taken to infinity while keeping the quantity
$$
\abs{\log q}=\frac{p^2}{2N}
$$
fixed. The resulting theory remains exactly solvable and provides an all-energy extension of the Schwarzian low-energy sector. Moreover, the gravitational dual of the double-scaled SYK model has recently been established at the disk level in terms of a dilaton model with a sine potential \cite{Blommaert:2024ymv,Blommaert:2024whf,Blommaert:2025avl}, providing a new and powerful perspective on these questions. Indeed, we show in this work that sine dilaton gravity can act as a similar gravitational bridge, connecting higher-dimensional black holes in de Sitter space and the double-scaled SYK model. 

More concretely, we will focus on Reissner-Nordström de Sitter black holes, a system with three distinct horizons: an inner horizon, an outer horizon, and a cosmological horizon. Extremal configurations arise when two or more of these horizons coincide and define the boundaries of the so-called sharkfin diagram, which encloses the physically allowed region in the $(M,Q)$ parameter space of the four-dimensional black hole. Configurations lying beyond these extremal boundaries are unphysical and correspond to the appearance of naked singularities in the four-dimensional geometry. Depending on which horizons approach each other, one obtains different near-horizon geometries and low-energy sectors: the cold, Nariai, and ultracold limits correspond to $\mathrm{AdS}_2 \times S^2$, $\mathrm{dS}_2 \times S^2$, and $\mathrm{Mink}_2 \times S^2$ near-horizon geometries respectively \cite{Romans:1991nq,Mann:1995vb,Booth:1998gf}.
The same near-horizon geometries, realized as fibered products over $\mathrm{AdS}_2$, $\mathrm{dS}_2$, and $\mathrm{Mink}_2$, also arise in extremal Kerr-de Sitter black holes, where the corresponding near-extremal gravitational dynamics is governed by JT gravity, as recently shown in \cite{Mariani:2025hee}.

\begin{figure}[h]
    \centering
    \includegraphics[width=0.9\linewidth]{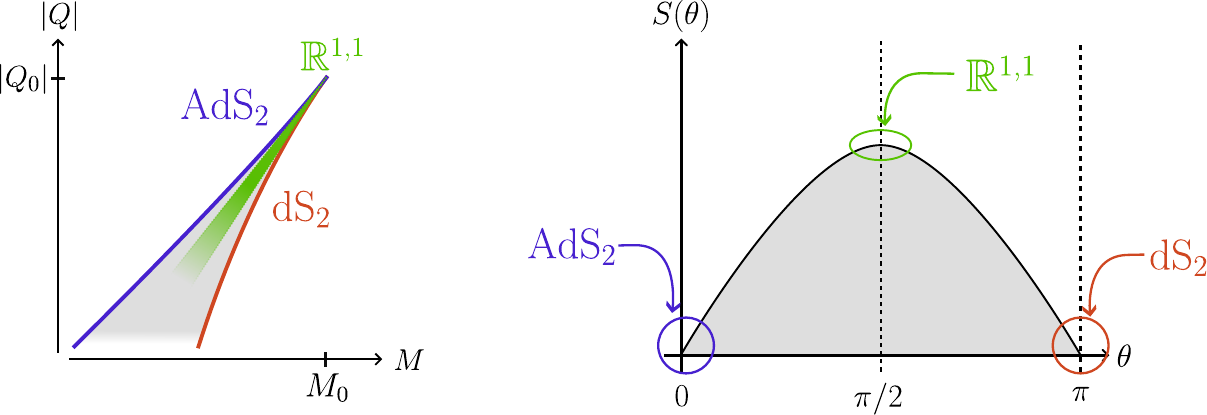}
    \caption{Analogy between the sharkfin diagram of $4\mathrm{d}$ RNdS$_4$ black hole, and the density of states of DSSYK and sine dilaton gravity.}
\label{fig:analogy}
\end{figure}

Remarkably, sine dilaton gravity interpolates between these three distinct two-dimensional JT models, with different regions of its phase space capturing the dynamics associated with each near-horizon geometry, as illustrated in Figure \ref{fig:analogy}. In particular, the lower edge of its spectrum describes the $\mathrm{AdS}_2$ regime, the upper edge corresponds to the $\mathrm{dS}_2$ regime, while the midpoint of the spectrum, characterized as containing the maximally entropic states, can be associated with flat spacetimes. 

Among these configurations, understanding the ultracold limit for which all three horizons coincide and the near-horizon geometry is flat, is particularly subtle since it is even more singular than the cold and Nariai regimes of the black hole leading to quantum effects not encountered in the other cases.

Our analysis of this model focuses on  understanding how the pathologies that arise in quantizing this model are to be treated. A naive treatment of the two-dimensional flat-space model governing fluctuations around the ultracold black hole leads to a Hagedorn growth of the density of states.
In particular, it predicts the existence of states that violate the physical bounds of the four-dimensional black hole (i.e. to stay within the sharkfin diagram). 
One may attempt to include the first subleading correction as a cosmic horizon regulator to the dilaton potential, however this still does not lead to a well-defined quantum theory. 
We propose instead to take inspiration from sine dilaton gravity by gauging a discrete shift symmetry in the reduced phase space of the model. This introduces a Gaussian suppression of sufficiently energetic states, effectively suppressing states corresponding to geometries with naked singularities.

Towards this goal we, in \textbf{section \ref{sec:RNdS}}, review some core properties of Reissner Nordstr\"om black holes in $3+1$d de Sitter space (RNdS$_4$), and define the ultracold and our near-ultracold reference system.
After that, we deduce the 2d dilaton gravity model that describes fluctuations around the near-ultracold black hole, as flat space JT gravity corrected with a quadratic term in the dilaton potential:
\begin{equation}
S_{2\mathrm{d}} = S_0 + \frac{1}{16\pi G_N}  \int d^2x \sqrt{-g} \left[\phi R + 2 - \left|\log q\right|\phi^2 \right],
\end{equation}
with the $2\mathrm{d}$ parameters related to the $4\mathrm{d}$ near-ultracold black hole parameters through \eqref{eq:2dNewton} and \eqref{eq:2d4ddictionary}, and the conversion scale \eqref{eq:aaa}. In order to understand the quantization of this model of fluctuations, in \textbf{section \ref{s:dilgravreview}} we review the salient features of $2\mathrm{d}$ dilaton gravity models and their quantization.
\textbf{Sections \ref{s:flatJT}, \ref{s:ref1}} and \textbf{\ref{s:ref2}} then implement this for our specific model at hand. This turns out to be subtle, and it will take us two refinements of the original flat space JT model, to land on something finite and physical.
We obtain the final partition function for the near-extremal fluctuations of the ultracold black hole as
\begin{equation}
Z(\beta) 
= \int_{-\infty}^{+\infty} dE e^{-\frac{64 G_4}{\delta^2_{Q^2}} E^2} e^{-\beta E},
\end{equation}
yielding a finite answer that is able to describe both positive and negative energy fluctuations. This result was written in \cite{Almheiri:2024xtw} as a toy model of quantum gravity and a limit of DSSYK. Our main interpretational result is that this model precisely describes the gravitational fluctuations of the near-extremal ultracold black hole, and at the same time provides a working example of flat space holography, a perspective we develop slightly by reinforcing the embedding within DSSYK in \textbf{section \ref{s:embed}}. We interpret our results physically in \textbf{section \ref{s:physconcl}}.
In \textbf{section \ref{s:concl}} we provide an outlook with various suggestive routes that we leave to the future. The appendix \ref{ssub:sharkfin2d4d} contains additional classical technical results on the sharkfin. 
\section{Near-Extremal Reissner-Nordstr\"om dS black holes}
\label{sec:RNdS}
We review the geometry of the RNdS$_4$ solution, define the ultracold and near-ultracold limit, and describe the model of fluctuations around the near-ultracold point we will develop in later sections.

\subsection{Classical geometry and near-ultracold point}
\begin{figure}[h]
    \centering
    \includegraphics[width=0.5\linewidth]{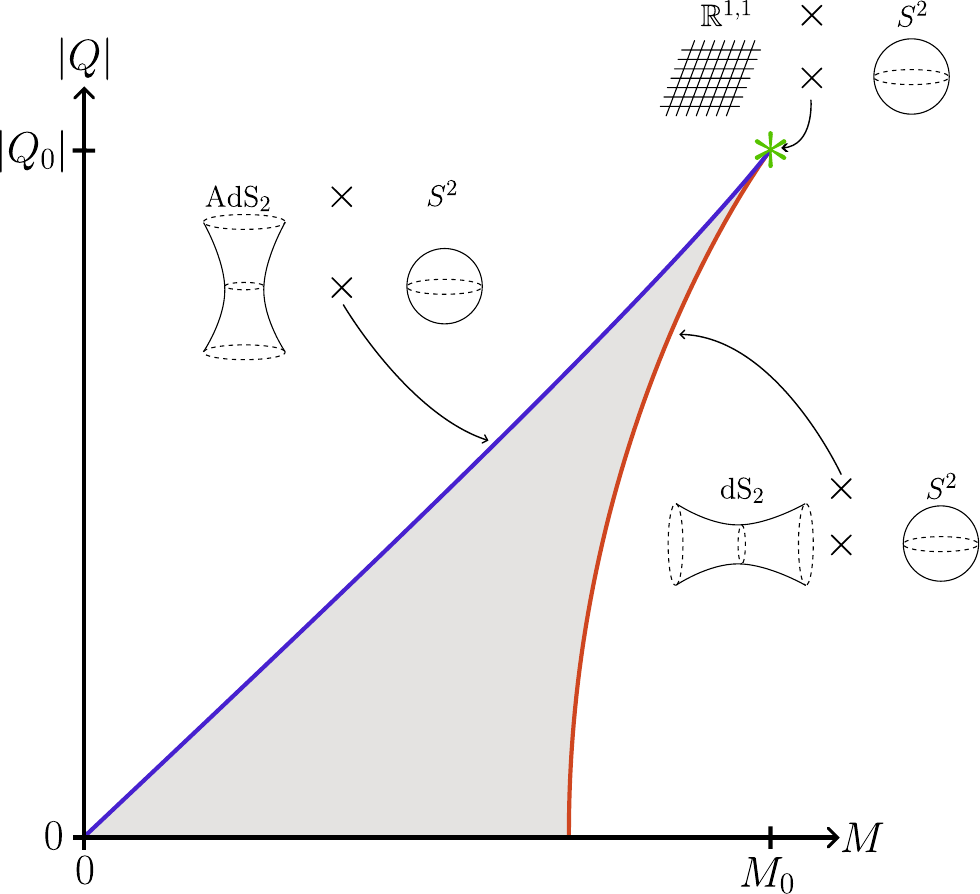}
    \caption{Sharkfin diagram for a RNdS$_4$ black hole for a fixed value of the cosmological constant $\Lambda_4$. Physical solutions lie inside the shaded area, while all the points outside of it correspond to naked singularities. The blue and the red lines correspond to the cold and the Nariai black holes respectively. The green star at the tip of diagram is the ultracold black hole.}
    \label{fig:sharkfin}
\end{figure}

Our starting point is the Reissner-Nordstr\"om black hole embedded in $3+1\mathrm{d}$ de Sitter spacetime. In the following, we will review some properties of this class of solutions, based on the original work of \cite{Romans:1991nq,Mann:1995vb,Booth:1997hq}. Starting from Einstein-Maxwell theory with a positive cosmological constant ($\Lambda_4>0$),
\begin{equation}
S_{\rm 4D}=\frac{1}{16\pi G_4}\int d^{4}x \sqrt{-g}\left({R}^{(4)}-2\Lambda_4 -F_{\mu\nu}F^{\mu\nu}\right)~ + \mathrm{bdy. \; terms}\label{eq:EML-action}
\end{equation}
where $\Lambda_4$ is related to the de Sitter radius via $\Lambda_4=3/\ell_\mathrm{dS}^2$ one finds
the classical saddles of this action to be black holes of mass $M$ and charge $Q$ specified via the metric and gauge field 
\begin{equation}
\begin{aligned}
	ds^{2}&=-V(r)d \tilde{t}^{2}+V^{-1}(r)d r^{2}+r^{2}(d \theta^2 +\sin^2\theta d\phi^2), \qquad A=  \frac{Q}{r}d \tilde{t}~,  \label{eq:dsrnds}
\end{aligned}
\end{equation}
in terms of the blackening factor of the metric
\begin{equation}
V(r) = 1-\frac{2M}{r}+\frac{Q^2}{r^2}-\frac{r^2}{\ell_\mathrm{dS}^2}  
= -\frac{(r- r_n)(r- r_-)(r - r_+)(r - r_c)}{r^2\ell_\mathrm{dS}^2}. 
\label{eq:wfds1}
\end{equation}
This blackening factor describes a class of charged black holes in de Sitter space with a (timelike) curvature singularity at $r=0$. For appropriate choices of $Q,M$ the solutions are classified in terms of the location of the three physical horizons in the geometry, i.e. the (real) roots of $V(r)$. They are denoted by $r_n < r_- < r_+ < r_c$ respectively; $r_\pm$ are the inner/outer black hole horizons, and $r_c$ is the cosmological horizon of de Sitter space. The fourth root of the polynomial $V(r)$, denoted $r_{n}$, is negative and therefore non-physical.
This yields relations of the set of roots and the labels of moduli space $M,Q,\ell_{\mathrm{dS}}$:
\begin{align}
\label{eq:coeffrel}
r_{n}&=-(r_-+r_+ +r_c),  &\ell_\mathrm{dS}^2 &= r_c^2 + r_+^2 + r_-^2  + r_c r_+ + r_+ r_- + r_-r_c,\\
\label{eq:mq}
    M &= \frac{(r_c + r_+)(r_+ + r_-)(r_- + r_c)}{2 \ell_\mathrm{dS}^2}, &Q^2 &= \frac{r_c r_+ r_- (r_c + r_+ + r_-)}{\ell_\mathrm{dS}^2}.
\end{align}
The presence of four real roots in $V(r)$ is assumed to avoid naked singularities. This imposes a constraint on moduli space. Namely a solution of the form \eqref{eq:wfds1} specified by a given choice of $(M,Q,\ell_{\text{dS}})$ is physical if it satisfies
\begin{equation}
\label{eq:4ddisc}
   \mathrm{Disc}_4 \equiv - 16 Q^6 + \ell_\mathrm{dS}^4\left(M^2 - Q^2\right) - \ell_{\mathrm{dS}}^2 \left( 27 M^4  -36 M^2 Q^2 -8Q^4\right) \geq 0,
\end{equation}
and $M \geq 0$.
For every fixed value of the de Sitter radius $\ell_\mathrm{dS}$ this restricts the space of classical solutions to what is colloquially referred to as the sharkfin, shown as the shaded region in Figure \ref{fig:sharkfin}.  
The boundaries of the sharkfin mark the occurrence of extremal solutions where two or more horizons merge \cite{Romans:1991nq, Booth:1997hq, Mann:1995vb}. Outside of the sharkfin is the unphysical region where naked singularities appear. One characterizing feature of the extremal geometries is the vanishing Hawking temperature at the extremal horizon $r_0$ since the degenerate root in the blackening factor guarantees\footnote{It is sometimes conventional to define the temperature as $V'(r_0)/4\pi$ leading to negative temperatures at the cosmological and inner black hole horizons \cite{Gibbons:1976ue, Banihashemi:2022htw}.}
\begin{equation}
\label{eq:THgen}
 T_\mathrm{Hawking}(r_0) =\frac{1}{4 \pi} \abs{\frac{d}{dr}V(r)}_{r = r_0} = 0.
\end{equation}
Above extremality, the Hawking temperatures of the three horizons are non-zero and given by
\begin{align}
\label{eq:THout}
T_\mathrm{Hawking}(r_+) = \frac{(r_c+2r_++r_-)(r_+-r_-)(r_c-r_+)}{4\pi r_+^2 \ell_\mathrm{dS}^2}, \\\
T_\mathrm{Hawking}(r_-) = \frac{(r_++2r_-+r_c)(r_+-r_-)(r_c-r_-)}{4\pi r_-^2 \ell_\mathrm{dS}^2},\\
T_\mathrm{Hawking}(r_c) = \frac{(r_++2r_c+r_-)(r_c-r_-)(r_c-r_+)}{4\pi r_c^2 \ell_\mathrm{dS}^2}.
\end{align}
Notice that, just like for the Schwarzschild de Sitter black hole, these temperatures are not equal and the full system is generally not in thermal equilibrium above extremality (see e.g. \cite{Johnson:2019ayc,Morvan:2022ybp,Morvan:2022aon} for some recent perspectives on this feature for the global system).\footnote{Except for the so-called lukewarm black holes which require $Q=M$. We will not be interested in this regime in this work.} 
We will later zoom in on physics just outside the outer horizon $r_+$ in a near-extremal regime, so that will be the thermal system we will focus on.

Merging the horizons leads to an enhancement of symmetry in the region near the outer black hole horizon, where the geometry typically factorizes in a direct product of the form $\mathcal{M}_2\times S^2$.
  For RNdS$_4$ black holes in $3+1$d this results in three distinct extremal configurations, corresponding to different regimes of the boundary of the sharkfin \ref{fig:sharkfin} and are coined respectively as:
\begin{itemize}
    \item The cold black hole, obtained when $r_-=r_+$, with an AdS$_2\times S^2$ near-horizon geometry;
    \item The Nariai black hole obtained when $r_+=r_c$, with a dS$_2\times S^2$ near-horizon geometry;
    \item The ultracold black hole, obtained when $r_-=r_+=r_c$, with a $\mathrm{Mink}_2\times S^2$ near-horizon geometry.
\end{itemize}

\begin{figure}[t!]
    \centering
\includegraphics[width=0.4\linewidth]{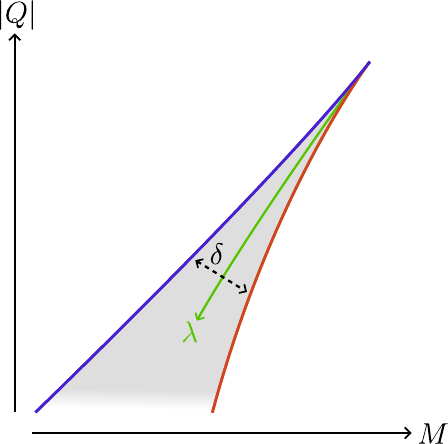}
     \caption{Zoom-in on the ultracold region. Moving inside the diagram along the $\lambda$ arrow is equivalent to following a near-extremal solution where the horizons are displaced symmetrically, i.e. $\delta=0$. The asymmetry in the horizon's displacement is implemented by moving along the dashed line corresponding to non-zero values of $\delta$. 
}
\label{fig:sharkfinld}
\end{figure}

\paragraph{The near-ultracold limit.}
In this work we will focus on the most constrained extremal solution of RNdS$_4$, known as the ultracold black hole. In the coordinates \eqref{eq:dsrnds} this corresponds to the point in parameter space where all three physical horizons coincide:
\begin{equation}
    r_+ = r_- = r_c \equiv r_0 = \frac{\ell_\mathrm{ds}}{\sqrt{6}}, 
\end{equation}
where the value of the ultracold radius $r_0$ is derived from \eqref{eq:coeffrel}. 
This limit describes the tip of the sharkfin where the cold and Nariai solutions intersect, and is characterized by the following values of mass and charge:
\begin{equation}
    Q_0^2 = \frac{r_0^2}{2}, \quad M_0 = \frac{2r_0}{3}.
\end{equation}
Since we are interested in studying the dynamics around the extremal ultracold regime, we will consider a near-extremal configuration, obtained by a small separation of the three horizons and a slight increase in the temperature.
Here it is useful to choose the following parametrization of the horizons 
\begin{align}
\label{eq:nearext}
   & r_+ - r_ - = (\lambda - \delta) ,\qquad r_+ - r_c = -(\lambda + \delta),    \\
      \Leftrightarrow \qquad &2\lambda = r_c-r_-, \qquad 2\delta = (r_c-r_+) - (r_+ -r_-) ,   
\end{align}
where $\lambda$ parametrizes symmetric splitting of the three roots while $\delta$ classifies the asymmetry in the separation of the horizons:
\begin{equation*}
\includegraphics[width=0.8\linewidth]{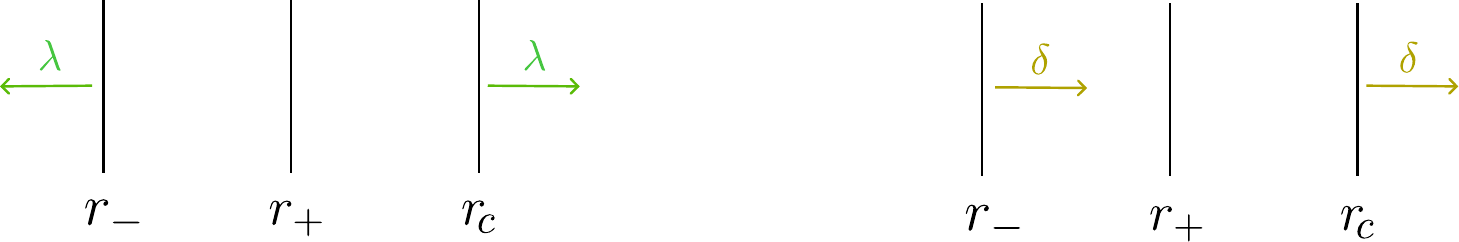} 
\vspace{-0.4cm}
\end{equation*}
Using the constraint $\ell^2_{\text{ds}}=6r_0^2$ in \eqref{eq:coeffrel}, and the above definitions of $\lambda$ and $\delta$, the outer horizon then reads
\begin{equation}
    r_+= -\frac{2}{3}\delta + r_0\sqrt{1-\frac{(\delta^2+3\lambda^2)}{18r_0^2}}.
    \label{eq:outerhorizon}
\end{equation}
For completeness, the expressions for the other two horizons are
\begin{align}
\label{eq:otherhorizon}
    r_-= \frac{1}{3}\delta -\lambda+ r_0\sqrt{1-\frac{(\delta^2+3\lambda^2)}{18r_0^2}}, \qquad
    r_c = \frac{1}{3}\delta +\lambda+ r_0\sqrt{1-\frac{(\delta^2+3\lambda^2)}{18r_0^2}}.
\end{align}
Rewriting the blackening factor yields 
\begin{equation}
    \label{eq:wfdsdelta}
    V(r) = -\frac{(r-r_+)((r - r_+ - \delta)^2 - \lambda^2)(r + 3r_+ + 2 \delta)}{\ell_\mathrm{dS}^2 r^2}. 
\end{equation}
While the constraint defining the sharkfin for fixed $\ell_{\text{dS}}$ \eqref{eq:4ddisc} simply results in the relations 
\begin{equation}
    \lambda \geq 0, \qquad \abs{\delta} \leq \lambda,
\end{equation}
imposing that the ordering of (real) roots is preserved.\footnote{For completeness the Nariai and cold black hole solutions corresponding to the boundary regions of the sharkfin are simply mapped to $\delta \to - \lambda$ and $\delta \to +\lambda$ respectively, with generic $\lambda$.}
In what follows, a \textit{near-ultracold} configuration is identified by a choice of $\lambda$, implementing a splitting between the horizons, followed by a choice of $\delta$, enforcing an asymmetric separation between $r_-$ and $r_c$ with the assumption that
\begin{equation}
\label{eq:defultracold}
  \abs{\delta} \ll \lambda \ll r_+.
\end{equation}
We will see that small $\lambda$ guarantees a near-extremal geometry and picking  asymmetric variations to be small $\abs{\delta} \ll 1$ guarantees the approximate flatness of the region near the outer black hole horizon $r_+$. 
In the sharkfin one can think of this as first tuning the ``distance" from the ultracold point by turning on $\lambda$ while $\delta$ then moves the solution from slightly positive (towards the Nariai limit) to slightly negative curvature (towards the cold limit), as illustrated in Figure~\ref{fig:sharkfinld}. 

Setting $\delta =0$ for now describes a codimension-1 curve into the sharkfin (the green curve in Figure~\ref{fig:sharkfinld}), we get upon series expansion of \eqref{eq:mq} in small $\lambda$ and using the explicit expressions \eqref{eq:outerhorizon}, \eqref{eq:otherhorizon} at $\delta=0$ that
\begin{align}
M \equiv M_0(1 + \delta_M) = M_0\big(1 - \frac{\lambda^2}{2r_0^2}\big) + \mathcal{O}(\lambda^4), \label{eq:M}\\
Q^2 \equiv Q_0^2(1+\delta_{Q^2}) = Q_0^2\big(1- \frac{4\lambda^2}{3r_0^2}\big) + \mathcal{O}(\lambda^4),
\label{eq:Q}
\end{align}
leading to the leading relation
\begin{equation}
\delta_M = \frac{3}{8} \delta_{Q^2}
\end{equation}
between the (dimensionless) deviations in the mass $M$ and charge $Q$ near the ultracold solution. For fixed small $\delta_{Q^2}$, this relation defines a single point very close to the ultracold tip (i.e. a point on the green curve in Figure~\ref{fig:sharkfinld}), which is our near-ultracold reference point around which we will consider quantum fluctuations in the later sections.\footnote{It is interesting to note that for fixed $\lambda$ varying $\delta$ from its extremal values $-\lambda$ to $\lambda$ results in variations of $Q$ and $M$ at order $\lambda^3$ suppressing  fluctuations in the "transverse" direction to the reference point.} 

On top of the mass and charge variations in \eqref{eq:M}, \eqref{eq:Q}, the symmetric separation of the horizons ($\delta=0$) in \eqref{eq:otherhorizon} gives rise to a non-vanishing Hawking temperature at the outer black horizon 
\begin{equation}
    T_\mathrm{Hawking}(r_+)=\frac{\lambda^2}{6\pi r_0^3} + \mathcal{O}\left(\frac{\lambda^4}{r_0^5}\right).
    \label{eq:Tplus}
\end{equation}

\paragraph{The infinite throat region in the ultracold.}
The simplifying feature of this limit is the claim that a divergent length flat throat is formed as one approaches the ultracold limit, and as such the geometry approximately factorizes as $\mathrm{Mink}_2\times S^2$, with the transverse 2-sphere of fixed size $r_0$ as one moves through the throat region. 
At first glance this claim may seem like a peculiar statement considering that in the standard coordinates \eqref{eq:wfds1} the ultracold limit corresponds to the case where an observer located outside of the black hole but inside the cosmological horizon is squeezed to a single point in the radial direction $r$, however this is just a sign of the coordinate system breaking down. 
Instead, one should study the proper distance $L_\mathrm{4d}$ between the outer and cosmological horizons: 
\begin{equation}
\label{eq:propdist4d}
    L_\mathrm{4d} = \ell_\mathrm{dS} \int_{r_+}^{r_c}\frac{rdr}{\sqrt{(r - r_-)(r - r_+)(r_c - r)(r - r_n)}} =  \ell_\mathrm{dS}\int_0^1 \frac{\frac{r_+}{r_c - r_+} + s}{\sqrt{s(1-s)(\frac{r_+ - r_-}{r_c -r_+} + s)(\frac{r_+ - r_n}{r_c - r_+}  + s)}} ds.
\end{equation} 
While the exact solution to this integral can be written in terms of elliptic integrals,\footnote{Using the  substitution $r= r_+ + (r_c - r_+)\frac{\sin^2 \phi}{1 + \frac{r_c - r_+}{r_+ - r_n} \cos^2\phi}$ and some algebraic manipulation of the resulting terms \cite{ByrdFriedman1954} one finds 
\begin{equation}
    L_{\mathrm{4d}} = \frac{2\ell_\mathrm{dS}}{\sqrt{(r_+ -r_-)(r_c -r_n)}}\left((r_+ - r_n)\Pi\left(\frac{r_c - r_+}{r_c -r_n};\sqrt{\frac{(r_c -r_+)(r_n -r_-)}{(r_+ - r_-)(r_c -r_n)}}\right) + r_n K\left(\sqrt{\frac{(r_c -r_+)(r_n -r_-)}{(r_+ - r_-)(r_c -r_n)}}\right)\right),
\end{equation}
in terms of the complete elliptic integrals of the first kind $K(k)$ and third kind $\Pi(\eta,k)$. 

} it is more insightful to directly expand the integrand in the ultracold limit \eqref{eq:nearext}:
\begin{align}
   L_\mathrm{4d} &\approx \frac{\ell_\mathrm{dS} r_+}{\lambda^{1/2}(4r_+ + 2\delta)^{1/2}(1 + \delta/\lambda)^{1/2}}\int_0^1 \frac{ds}{\sqrt{s(1-s)(s + \frac{\lambda -\delta}{\lambda + \delta}})}.
    \label{eq:L4d}
\end{align}
The integral itself is finite and one indeed finds
that as expected $L_\mathrm{4d}$ is divergent as $\lambda \to 0$ since to leading order in the ultracold regime \eqref{eq:defultracold}:
\begin{equation}
    L_\mathrm{4d} \approx \left(\frac{3}{2}\right)^{1/2}\frac{r_0^{3/2}}{\lambda^{1/2}}\sqrt{2}K\left(\frac{\sqrt{2}}{2}\right),
    \label{eq:dist4dleading}
\end{equation} 
with $K(k)$ the complete elliptic integral of the first kind. 
For completeness, one can of course perform an identical analysis of the other extremal limits of the black hole. In the cold limit $(r_+ - r_- = \epsilon)$, 
\begin{align}
    L_{\mathrm{cold}} 
    =&\simeq \ell_\mathrm{dS} \frac{r_+}{\sqrt{(-r_n + r_+)(r_c - r_+)}}\int_0^1 \frac{ ds}{\sqrt{s(s + \frac{\epsilon}{r_c - r_+})(1 - s)}} \sim \ln (\eps/(r_c -r_n)) \to \infty.
\end{align}
This is just AdS-JT gravity well known to also have divergent proper distance to the horizon due to the emergence of an infinite AdS$_2$ throat region close to the black hole horizon ($s \to 0$). 
In the Nariai black hole $r_c - r_+ =\epsilon \to 0$ instead the integration range also vanishes in the limit leading to a finite result
\textbf{\begin{align}
    L_{\mathrm{Nariai}} &= \ell_\mathrm{dS}\int_{0}^{\epsilon} \frac{(r + r_+)dr}{\sqrt{r(r + \epsilon )((r_c - r_-) - r)(r - (r_n - r_+)})} \sim \ell_\mathrm{dS}\frac{\pi r_+}{\sqrt{(r_+ - r_n)(r_+ - r_-)}}.
\end{align}}

\begin{figure}
    \centering
    \includegraphics[width=0.99\linewidth]{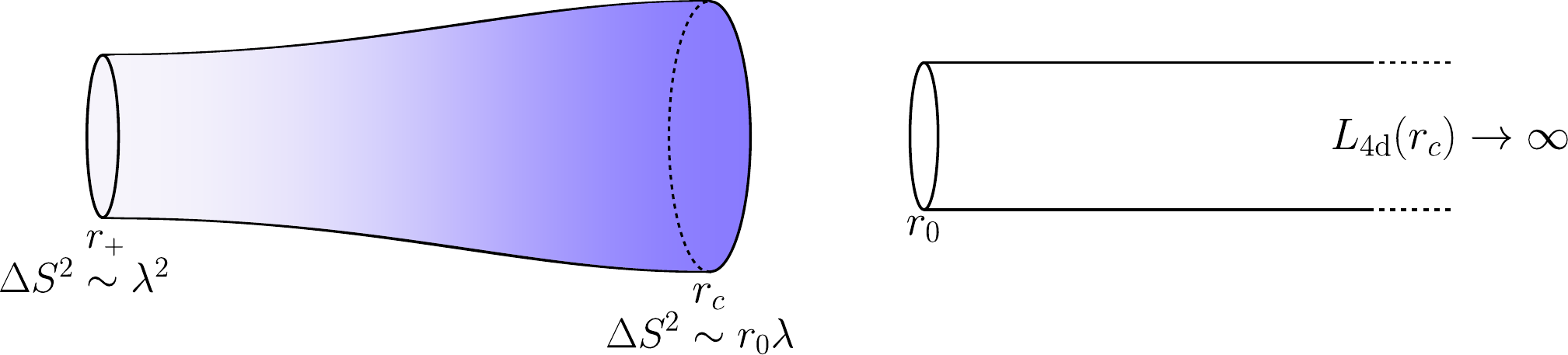}
    \caption{Size of the radius of the transverse 2-sphere $S^2$ as function of proper distance to the outer horizon using the inverse of \eqref{eq:propdist4d} with coloring proportional to the exact curvature $R_{2d}(L_{4\mathrm{d}})$. The near-ultracold limit (left) shows development of a nearly flat region in the (outer) near-horizon regime. Changes in area of the transverse 2-sphere at $r_+$ are suppressed by a factor of $\lambda/r_0$ with respect to the cosmological horizon $r_c$. In the strict ultracold limit (right) the cosmological horizon is pushed to infinite proper distance, while the $2\mathrm{d}$ curvature $R_{2d}$ vanishes. The size of the $S^2$ is constant at $4\pi r_0^2$ throughout the region.}
    \label{fig:geom_uc}
\end{figure}

Returning to the ultracold black hole, we found that indeed the proper distance between horizons is divergent reflecting the formation of an infinite
 throat. The flatness of the near-horizon region can be seen at the level of the $2$d Ricci scalar. Taking into account only the $(t,r)$ sector of the metric,
\begin{equation}
    g_{ab}dx^adx^b=-V(r)dt^2+\frac{dr^2}{V(r)},
\end{equation}
and considering the near-extremal configuration obtained through \eqref{eq:nearext}, the $2$d Ricci scalar in the near-horizon region reads
\begin{equation}
    R_{\mathrm{2d}}= -V''(r(x)) = \frac{4x}{r_0^3} + \mathcal{O}\left(\frac{\lambda^2}{r_0^4},\frac{x^2}{r_0^4}\right), 
    \label{eq:2dRicci}
\end{equation}
where $x$ is the $2$d radial coordinate from the horizon, $r\rightarrow r_0+x$ and thus bounded by $\abs{x} \lesssim \lambda$.
If we evaluate \eqref{eq:2dRicci} exactly at the outer black hole horizon, $r_+$ \eqref{eq:outerhorizon}, we find to leading order 
\begin{equation}
    R_{\mathrm{2d}}\big|_{r_+}\approx -\frac{8\delta}{3r_0^3} + \frac{7\lambda^2}{3r_0^4},
    \label{eq:R4d}
\end{equation}
which for $\abs{\delta} \ll \lambda \ll r_0$ ensures approximate flatness of the near-horizon region. In the strict ultracold case, corresponding to $\delta= \lambda = 0 $, the $2$d near horizon geometry is exactly flat. 
In contrast the curvature at the inner black hole horizon and cosmological horizon are respectively
\begin{equation}
    R_{2\mathrm{d}}\vert_{r_-} \approx  -\frac{4\lambda}{r_0^3}, \qquad     R_{2\mathrm{d}}\vert_{r_c}  \approx  \frac{4\lambda}{r_0^3}.
\label{eq:Rcm}
\end{equation}
More generally we thus find that in the specified limit the curvature in the near-horizon regime, $\abs{x}/\lambda \ll 1$,  is suppressed with respect to the other relevant curvature scales in the setup. Especially the curvature close to the outer black hole horizon is smaller by an order of $\lambda/r_0$ with respect to the curvature at the other horizons in the geometry. 
Summarizing, the $(t,r)$-geometry in the near-ultracold black hole, shown in Figure~\ref{fig:geom_uc}, forms a nearly flat infinitely long throat regulated by the inner and cosmological horizons respectively.   

\subsection{Dimensional reduction of RNdS$_4$}
\label{sub:dimred}
In any of the near-extremal regimes of RNdS$_4$ black holes presented above, the geometry does not factorize exactly but rather develops a finite (but large) throat where the geometry is approximately factorized. This is the limit where the dynamics of the model is dominated by the $s$-wave sector of perturbations and it is exactly this limit that leads to the highly successful description of near-extremal Reissner-Nordstr\"om black holes in $4\mathrm{d}$ flat space by JT-gravity \cite{Nayak:2018qej,Larsen:2018iou,Castro:2021wzn}. In this subsection, we are interested in finding an effective theory describing excitations above extremality for ultracold black holes, building in part on \cite{Castro:2022cuo}. 

 Starting with the $4\mathrm{d}$ Einstein-Maxwell action \eqref{eq:EML-action} 
\begin{equation}
S_{\rm 4d}=\frac{1}{16\pi G_4}\int_{{\mathcal{M}}} d^{4}x \sqrt{-g}\left(R^{(4)}-2\Lambda_4 -F_{\mu\nu}F^{\mu\nu}\right),
\end{equation}
we perform a dimensional reduction of this theory via the following spherically symmetric ansatz for the metric:   
\begin{equation}
\label{eq:metric4d}
\begin{aligned}
 ds^2_4&=g^{(4)}_{\mu\nu} d x^\mu d x^\nu= \frac{\Phi_0}{\Phi} g_{ab} d x^a d x^b + \Phi^2 \left(d\theta^2+\sin^2\theta d\phi^2\right)~,\\
  F&= F_{ab} d x^a \wedge d x^b.   
\end{aligned}
\end{equation}
Here $g_{ab}$ is a $2$d metric, depending only on the 2d coordinates $x^a$, $a,b=(0,1)$, $\Phi(x)$ is the dilaton scalar field and $\Phi_0$ is a constant. 
We interpret $\Phi(x)$ as a (spherically symmetric) fluctuation of the radial coordinate, and the constant $\Phi_0$ is simply the ultracold radius, $\Phi_0=r_0$.
Upon dimensional reduction of the Einstein–Maxwell action \eqref{eq:EML-action} with respect to \eqref{eq:metric4d}, we obtain the effective two-dimensional action\cite{Castro:2022cuo}:
\begin{equation}\label{eq:2daction}
\begin{aligned}
{S_{\rm 2d}=\frac{1}{4 G_4} \int_{{\mathcal{M}_2}} d^2x \sqrt{-g}\Phi^2 \left( {R}+ 2\frac{\Phi_0}{\Phi^3}-{2\Lambda_4}\frac{\Phi_0}{\Phi}-\frac{\Phi}{\Phi_0} F_{ab}F^{ab} \right).}
\end{aligned}
\end{equation}
The $4$d Newton's constant $G_4$ is effectively related to the $2$d one by a factor of
$4\pi$, the volume of the transverse $2$-sphere, and $R$ is the two-dimensional Ricci scalar associated to the 2d metric $g_{ab}$. So far we have not specified the boundary term of this action. A natural choice is given by Einstein-Maxwell theory in a fixed charge ensemble. This is achieved by supplementing the action \eqref{eq:EML-action} by the $3+1$d boundary term 
\begin{equation}
    S_{b}^{\text{em}}=\frac{1}{4\pi G_4}\int_{\partial\mathcal{M}} d^3 x\sqrt{-h}\hat{n}_{\mu}F^{\mu\nu}A_{\nu}.
    \label{eq:boundary}
\end{equation}
This boundary term \cite{Coleman:1991ku} has previously been argued for e.g. in the setting of black hole pair creation \cite{Hawking:1995ap} and complexity \cite{Brown:2018bms} making it a natural choice in our setting.
Dimensionally reducing \eqref{eq:boundary} leads to a $1$d boundary term that can be rewritten in the form of a $1+1$d bulk by applying the divergence theorem \cite{Brown:2018bms}:
\begin{equation}
    S_{\text{bulk,2d}}^{\text{em}}=\frac{1}{2G_4}\int_{{\mathcal{M}_2}} d^2x \sqrt{-g}\frac{\Phi^3}{\Phi_0}F_{ab}F^{ab}.
    \label{eq:bulkterm}
\end{equation}

As a boundary term this does not change the equations of motion, i.e. Maxwell equations, but generically contributes to the dilaton potential and especially the on-shell value of the action \cite{Brown:2018bms}.
Variation of the Maxwell term plus the boundary term 
\eqref{eq:boundary}, leads to Maxwell equations with boundary condition
\begin{equation}
    \nabla^{\mu}F_{\mu\nu}=0, \qquad \delta(n_{\mu}F^{\mu\nu})\vert_{\partial \mathcal{M}}=0,
\end{equation}
where the last equation simply means that the total charge $Q$
is fixed. The field strength is obtained as a solution to Maxwell's equations and reads
\begin{equation}
F_{ab}= Q \frac{\Phi_0}{ \Phi^3}\sqrt{-g}\epsilon_{ab},\quad F^2=-2Q^2\frac{\Phi_0^2}{\Phi^6},
\label{eq:maxwellsol}
\end{equation}
the electric charge $Q$ here is a constant, and $\epsilon_{ab}$ is the $2$d Levi-Civita tensor.

In the following, we will study the system after a thermal near-extremal perturbation around the ultracold background. This amounts to a perturbation of the charge of the black hole
\begin{equation}
    Q^2\rightarrow Q_0^2(1 +\delta _{Q^2}) = Q_0^2(1 - \abs{\delta_{Q^2}}),
    \label{eq:Qtransf}
\end{equation}
where $Q_0$ is the charge of the extremal solution, $Q_0^2=\Phi_0^2/2$, and $\delta_{Q^2}$ is an infinitesimal variation of the (squared) charge. The fluctuation naturally satisfies $\delta_{Q^2} < 0$ to ensure the heated black hole to satisfy the physical bound \eqref{eq:4ddisc}. 
At the level of the sharkfin diagram, this corresponds to moving diagonally inside the region of allowed physical solutions starting from the ultracold point at the tip of the diagram (Figure~\ref{fig:sharkfinld}), corresponding to a value of the constant $\Phi_0$ of $\Phi_0=\ell_\mathrm{ds}/\sqrt{6}$, varying both the charge and the mass of the black hole by displacing the horizons. 

With the near-horizon limit in mind, we implement a field redefinition of the dilaton (or equivalently, since $\Phi$ is the radial coordinate, a coordinate transformation on the classical solution) 
\begin{equation}
\label{eq:fieldredef}
    \Phi = \Phi_0 \sqrt{1 + \phi_0 + \alpha\phi}, 
\end{equation}
where $\phi_0$ and $\alpha$ are generic parameters to be determined shortly. The quantity $\phi_0$ characterizes a zero-mode offset of the extremal radius; moving it to a near-ultracold regime. The parameter $\alpha$ sets the scale for the radial (dilaton) fluctuations around this near-ultracold reference point. 
In this parametrization, and integrating out the Maxwell term, the action takes the form 
\begin{align}
    S_\mathrm{2d} = &\frac{1}{4G_4}\int_{{\mathcal{M}_2}} d^2x \sqrt{-g} R \,\Phi_0^2(1 + \phi_0) \nonumber \\
    &+ \frac{\Phi_0^2\alpha}{4 G_4 }\int _{{\mathcal{M}_2}} d^2x\sqrt{-g}\left[ \phi R + \frac{1 - \phi_0^2  - 2Q^2/\Phi_0^2 - 2\alpha \phi_0 \phi - \alpha^2\phi^2}{\Phi_0^2 \alpha (1 + \phi_0 + \alpha\phi)^{3/2}}\right].
    \label{eq:I2dexact}
\end{align}
Note that the first term is purely topological and proportional to the classical black hole entropy  
\begin{equation}
  S_0=  \frac{\Phi_0^2(1 + \phi_0)}{4G_4} \int_{{\mathcal{M}_2}} d^2x \sqrt{-g}R =  \frac{4\pi \Phi_0^2(1 + \phi_0)}{4G_4}\chi({\mathcal{M}_2}),
\end{equation}
in terms of the Euler characteristic $\chi$ (upon Wick rotation). Since $\Phi_0 = r_0$ this is indeed proportional to the area of the ($4\mathrm{d}$) horizon with $\phi_0$ acting as a correction to the size of the transverse sphere as the system is perturbed away from the ultracold point.

Note that so far all transformations have been exact.
At this point, the action in \eqref{eq:I2dexact} describes a specific 1+1d dilaton gravity model of the form
\begin{equation}
    S= \frac{1}{16\pi G_N}\int_{{\mathcal{M}_2}} d^2x\sqrt{-g}(\phi R+V(\phi)),
    \label{eq:genericdilaton}
\end{equation}
where $V(\phi)$ is the dilaton potential. In the strict ultracold regime, at zero-th order leading approximation, the dilaton potential in \eqref{eq:I2dexact} reduces to 
\begin{equation}
V(\phi) = \frac{1-2Q_0^2/\Phi_0^2}{\Phi_0^2} \overset{!}{=} 0,
\end{equation}
and the model reduces to flat space JT gravity without cosmological constant:
\begin{equation}
\frac{1}{16\pi G_N} \int_{{\mathcal{M}_2}} d^2x \sqrt{-g} \,\phi R.
\end{equation}
The classical solution of this model describes a zero-temperature black hole, matching with the strict ultracold black hole in four dimensions.

The near-ultracold limit is implemented by expanding for 
\begin{equation}
\delta_{Q^2} \ll 1, \quad \phi_0, \,\,\alpha \phi \ll 1.
\label{eq:expansions}
\end{equation} 
Here both $\delta_{Q^2}$ and $\phi_0 + \alpha \phi$ are treated as infinitesimal expansion parameters at the same perturbative order. We expand the action \eqref{eq:I2dexact} up to quadratic order in these fluctuations.\footnote{Note that this expansion only assumes being close to the tip of the sharkfin, however it does not a priori limit to the case $\abs{\delta} \ll \lambda$ i.e the flat region. One way to appreciate this is the fact that tuning the curvature within the sharkfin requires fixing the relation between $\delta_{Q^2}$ and $\delta_M \equiv (M/M_0) -1 $, however the bulk action only depends on $\delta_{Q^2}$ explicitly. We comment more on this in appendix \ref{ssub:sharkfin2d4d}.} In this limit, the action takes the form
\begin{equation}
\label{eq:newdilaction}
     S_\mathrm{2d} = S_0 + \frac{\Phi_0^2\alpha}{4 G_4} \int_{{\mathcal{M}_2}} d^2x \sqrt{-g} \left[\phi R - \frac{1}{\Phi_0^2\alpha}\left(\delta_{Q^2}(1 -\frac{3}{2} \phi_0) + \phi_0^2 +  2\phi \alpha\left(\phi_0 - \frac{3}{4}\delta_{Q^2}\right)   + \alpha^2\phi^2\right)\right]. 
\end{equation}
The action takes a particularly suggestive form if now set
\begin{equation}
    \phi_0 = \frac{3\delta_{Q^2}}{4}.
    \label{eq:phi0}
\end{equation}
The choice of $\phi_0$ implements an expansion around the near-ultracold black hole of radius $\Phi_0\sqrt{1+\frac{3\delta_{Q^2}}{4}}$, with radial fluctuations described by $\alpha \phi$. Uniquely for this choice, one eliminates the term linear in the dilaton potential in \eqref{eq:newdilaction}. This corresponds to an expansion around the extremum of the dilaton potential, and picks for us a unique route down the sharkfin starting in the ultracold tip (see Figure \ref{fig:sharkfinld}). Ultimately, this is just a gauge choice for the origin of $\phi$ around which to expand, and thus is not physical. The main physical feature is that we retain the quadratic term in the potential at the same perturbative level. 

Let us first focus on the constant term in the dilaton potential of \eqref{eq:newdilaction}. Together with the $\phi R$ term, this leads to a flat JT model of the type
\begin{equation}
\frac{1}{16\pi G_N} \int_{{\mathcal{M}_2}}  d^2x \sqrt{-g} \left[\phi R - \Lambda\right],
\end{equation}
with $2\mathrm{d}$ Newton constant 
\begin{equation}
\label{eq:2dNewton}
G_N = \frac{G_4}{4\pi\Phi_0^2\alpha},
\end{equation}
and negative cosmological constant
\begin{equation}
\Lambda=\frac{1}{\Phi_0^2\alpha}\left(\delta_{Q^2} - \frac{9}{16}\delta_{Q^2}^2\right)<0.
\end{equation}
Choosing the scale $\alpha$ in the parametrization \eqref{eq:fieldredef} as
\begin{equation}
\label{eq:alpha}
\alpha = -\frac{\left(\delta_{Q^2} - \frac{9}{16}\delta_{Q^2}^2\right)}{2\Phi_0^2} >0,
\end{equation}
we can rewrite this in a canonical form
\begin{equation}
\frac{1}{16\pi G_N} \int d^2x \sqrt{-g} \left[\phi R + 2\right].
\end{equation}
With this choice of scale $\alpha$, the full $2\mathrm{d}$ action becomes\footnote{This action was recently encountered in \cite{Chen:2025jqm}.}
\begin{equation}
\frac{1}{16\pi G_N} \int_{{\mathcal{M}_2}} d^2x \sqrt{-g} \left[\phi R +2- \left|\log q\right|\phi^2 \right],
\end{equation}
where\footnote{We denote this parameter as $\abs{\log q}$ with the foresight of the connection of this model with DSSYK in section \ref{s:embed}.  }
\begin{equation}\label{eq:2d4ddictionary}
\left|\log q\right| = \frac{\alpha}{\Phi_0^2} = -\frac{\left(\delta_{Q^2} - \frac{9}{16}\delta_{Q^2}^2\right)}{2\Phi_0^4}
\end{equation}
and the near-ultracold assumption $\lambda \ll r_0 = \Phi_0$ is encoded in the fact that $\abs{\log q} \ll 1$.
These choices eliminate the term linear in the dilaton and rescale the constant term in the dilaton potential to be $\delta_{Q^2}$ independent. Notice that the choice $\phi_0 = 3\delta_{Q^2}/4$ is consistent with the expansion we took in \eqref{eq:expansions}, where $\phi_0$ and $\delta_{Q^2}$ are of the same perturbative order. 

The final action takes the form
\begin{equation}
\boxed{
    S_{2\mathrm{d}} = S_0 + \frac{1}{16\pi G_N}  \int_{{\mathcal{M}_2}} d^2x \sqrt{-g} \left[\phi R + 2 - \left|\log q\right|\phi^2 \right]}, \qquad \frac{1}{16\pi G_N} = \frac{\Phi_0^4 \abs{\log q}}{4G_4}.
    \label{eq:2dgaugefixed}
\end{equation}
Note that due to $\abs{\log q} \ll 1$, $G_N$ is large and the resulting $2\mathrm{d}$ theory is strongly coupled in the near-ultracold regime of interest in this work.

At leading order in the fluctuations the action \eqref{eq:2dgaugefixed} leads to a vanishing two-dimensional Ricci scalar, and is therefore consistent with flat JT gravity. The first subleading correction then serves as a regulator introducing the effects of the inner and cosmological horizons on the system. Much as JT gravity plays a central role in describing the near-horizon dynamics of near-extremal Reissner-Nordstr\"om black holes and in capturing their quantum corrected low temperature behavior, we expect that the two-dimensional model \eqref{eq:2dgaugefixed} and its quantization can describe the low-temperature quantum behavior of the Reissner-Nordstr\"om de Sitter black hole in the near-ultracold regime.

\section{Review: Dilaton gravity technology}
\label{s:dilgravreview}
Our task is now to understand quantum fluctuations in this regime, described by \eqref{eq:2dgaugefixed}, and hence to perform the Euclidean gravitational path integral
\begin{equation}
\label{eq:tocomp}
Z(\beta) \overset{?}{\sim} \int [\mathcal{D}g_{\mu\nu}][\mathcal{D}\phi]e^{ \frac{1}{16\pi G_N}  \int d^2x \sqrt{g} \left[\phi R + 2 - \left|\log q\right|\phi^2 \right] - S_{\text{bdy}}},
\end{equation}
on the disk topology with circumference $\beta=T^{-1}$, and with $G_N$ related to the $4\mathrm{d}$ parameters through \eqref{eq:2dNewton}. Since we restrict to the lowest topology, we can drop the extremal term $S_0$ in \eqref{eq:2dgaugefixed}. We will make some comments on higher topology throughout this work, but this is not our focus here. The temperature $T$ is to be thought of as the temperature at the outer horizon $r_+$ of the near-ultracold black hole, as we will elaborate on below. 

In order to facilitate such a computation, we must specify our starting point in more detail. In particular, we need to detail the boundary action $S_{\text{bdy}}$ and pinpoint how we are going to quantize this model \eqref{eq:tocomp}. The Lorentzian signature action of general dilaton gravity with dilaton potential $V(\Phi)$ takes the form
\begin{equation}\label{eq:dila1}
    S = \frac{1}{16\pi G_N} \int_{\mathcal{M}} \mathrm{d}^2 x \sqrt{-g}\,\bigl(\phi R + V(\phi)\bigr)
      + \frac{1}{8\pi G_N}\oint_{\partial \mathcal{M}} \mathrm{d}t \sqrt{-h}\left(
        \phi_\partial K - \sqrt{\int_{c}^{\phi_{\partial}} V(\phi)\, \mathrm{d}\phi}
      \right).
\end{equation}
We have included here the boundary term action $S_{\text{bdy}}$, consisting of the usual Gibbons-Hawking-York boundary term and a holographic counterterm (see e.g. \cite{Belaey:2025kiu}), with Dirichlet boundary condition for the dilaton $\phi_\partial$ fixed to some large but finite value. The latter is the natural generalization of the one used in asymptotic AdS holography and we will stick with it for this work. When there are cosmological horizons in the game, the physical location of the actual boundary $\partial \mathcal{M}$ is more subtle, and we come back to it in section \ref{s:embed}. 

We now collect and review the relevant technology for exact treatments of dilaton gravity, both classically and quantum mechanically. We summarize three features of dilaton gravity models that we will leverage to deduce their physical features. These will be applied in the next main sections to decide on how to properly quantize the dominant near-ultracold gravitational fluctuations.

\subsection{Feature 1: Classical solution and thermodynamics}
In the Schwarzschild gauge of the model, the metric and dilaton solution can be written as \cite{Gegenberg:1994pv,Banks:1990mk,Witten:2020ert}
\begin{equation}
\label{eq:dilaa}
    \mathrm{d}s^2 =
    -\frac{1}{a^2}\left(W(r)-W(r_h)\right)\mathrm{d}t^2
    + a^2\frac{\mathrm{d}r^2}{W(r)-W(r_h)} ,
    \quad W(\phi(r)) \equiv  \int^\phi V(\phi)\,\mathrm{d}\phi,
\end{equation}
with $\phi = ar+b$ and where we defined the prepotential $W(\phi)$ as the integral of $V(\phi)$. This is a black hole geometry with horizon at $r = r_h$. The constant $b$ can be set to zero by a choice of origin. The coefficient $a$ however contains physics, since it determines the slope of the dilaton at the holographic boundary at $\abs{r} \to \infty$, and boundary diffeomorphisms are physical, rescaling the energy and temperature of configurations.\footnote{We have deliberately used notation to allow the boundary to be at \emph{complex} infinity, such that $\abs{r} \to \infty$, as will happen in our case and in sine dilaton gravity.} The thermodynamic properties of the general dilaton gravity black hole solution \eqref{eq:dilaa} are given by \cite{Mertens:2022irh} ($\phi_h=ar_h$):
\begin{equation}\label{eq:thermo}
        \beta = \frac{4\pi}{V(\phi_h)}a ,
    \qquad
    S = \frac{\phi_h}{4G_N} ,
    \qquad
    E_{\mathrm{ADM}} = \frac{1}{16\pi G_N a} W(\phi_h) .
\end{equation}
The quantity $a$ sets the scale of all dimensionful quantities. Since the boundary condition is
\begin{equation}
\phi_{\partial} \equiv \phi(\abs{r}\to\infty) \approx ar,
\end{equation}
the information on $a$ is encoded in the boundary term of the dilaton gravity action \eqref{eq:dila1}, and is a property of the model (and not just of a specific classical solution). One way to fix it in our model, is to compare these thermodynamic BH equations \eqref{eq:thermo} to those of the $4\mathrm{d}$ near-ultracold black hole, as we will do in the next section.

It is sometimes more convenient to set $a=1$ and work with dimensionless quantities:
\begin{equation}\label{eq:dila}
    \mathrm{d}s^2 =
    -\left(W(\phi)-W(\phi_h)\right)\mathrm{d}t^2
    + \frac{\mathrm{d}\phi^2}{W(\phi)-W(\phi_h)} ,
    \qquad W(\phi) \equiv \int^\phi V(y)\,\mathrm{d}y.
\end{equation}
The radial coordinate is now directly the dilaton $\phi$, with the horizon at $\phi = \phi_h$. We can then reinstate the units at the end of a calculation by inserting factors of $a$ where appropriate.

\subsection{Feature 2: Two-sided phase space description}
Our aim is to find a two-sided boundary Hamiltonian for general dilaton gravity, in the spirit of \cite{Harlow:2018tqv}. However, the region at $\phi \rightarrow +\infty$ in the geometry \eqref{eq:dila} is not a standard AdS$_2$ boundary. As a consequence, for generic potentials there is not a unique geodesic connecting two arbitrary boundary points, unlike in AdS$_2$.
This issue was resolved in \cite{Blommaert:2024whf} by introducing an effective geometry obtained from the original geometry \eqref{eq:dila} through the conformal mapping
\begin{equation}
\label{eq:conf_mapping}
    \mathrm{d}s^2 = e^{2\omega}\,\mathrm{d}s^2_{\mathrm{eff}} ,
    \qquad
    \mathrm{d}s^2_{\mathrm{eff}}
    = -\left(\rho^2 - \rho_h(\phi_h)^2\right)\mathrm{d}t^2
    + \frac{\mathrm{d}\rho^2}{\rho^2 - \rho_h(\phi_h)^2} ,
\end{equation}
where $\mathrm{d}s^2_{\mathrm{eff}}$ represents a standard AdS$_2$ black hole with horizon at $\rho=\rho_h$. Imposing the conformal relation \eqref{eq:conf_mapping} requires
\begin{equation}\label{eq:PhiPrime}
    \phi'(\rho)
    = e^{2\omega}
    = \frac{W(\phi) - W(\phi_h)}{\rho^2 - \rho_h^2} .
\end{equation}
This can be viewed as a first-order differential equation that relates the original radial coordinate $\phi$ to the new one $\rho$. To determine the function $\rho_h = \rho_h(\phi_h)$, we evaluate the above expression near $\rho=\rho_h$. Expanding $W(\phi)$ around the horizon yields
\begin{equation}
    \phi'(\rho_h)
    =
    \frac{V(\phi_h)\left[\phi(\rho_h) + \phi'(\rho_h)(\rho-\rho_h) - \phi_h\right]}
    {2 \rho_h (\rho-\rho_h)} .
\end{equation}
Requiring $\phi(\rho_h) = \phi_h$, which ensures that the horizon area and therefore the entropy remain invariant under the conformal mapping, implies
\begin{equation}
   2 \rho_h = V(\phi_h) .
\end{equation}

Smoothness at the black hole horizon in the effective geometry \eqref{eq:conf_mapping} further implies that the effective inverse Hawking temperature is
\begin{equation}
    \beta_{\mathrm{eff}} = \frac{2\pi}{\rho_h}= \frac{4\pi}{V(\phi_h)} .
\end{equation}
Since both the temperature and entropy coincide with those of the original geometry, the first law of thermodynamics guarantees that all thermodynamic quantities \eqref{eq:thermo} remain unchanged under the conformal transformation.

As a field theory, the phase space of any dilaton gravity is infinite dimensional. Working instead with physical (= diff-invariant) observables, the (reduced) classical phase space of the effective AdS$_2$ model is two-dimensional and is characterized by the symplectic form \cite{Harlow:2018tqv}
\begin{equation}
    \omega = \mathrm{d}T \wedge \mathrm{d}E ,
\end{equation}
where $T$ denotes the two-sided Lorentzian time on the effective AdS$_2$ geometry, and $E$ is its conjugate ADM energy given in \eqref{eq:thermo}:
\begin{equation}
    \begin{tikzpicture}[baseline={([yshift=-.5ex]current bounding box.center)}, scale=0.7]
  \pgftext{\includegraphics[scale=1]{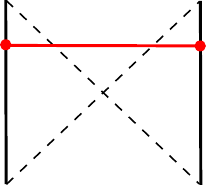}}
 at (0,0);
    \draw (2.4,0.7) node {$T/2$};
    \draw (0,1.2) node {${\color{red}L}$};
  \end{tikzpicture}
\end{equation}

It is now convenient to perform a canonical transformation and introduce new phase-space variables: the length $L$ of the Einstein-Rosen bridge connecting the two asymptotic boundaries of the effective AdS$_2$ geometry \eqref{eq:conf_mapping} and its conjugate momentum $P$. The bridge length as a function of temperature is known \cite{Harlow:2018tqv} and in the present case takes the form
\begin{equation}\label{eq:Lor-rel}
    L
    = 2 \log\!\left(
        \frac{\beta_{\mathrm{eff}} \cosh\!\left(\frac{\pi T}{\beta_{\mathrm{eff}}}\right)}{2\pi}
      \right)
    = 2 \log\!\left(
        \frac{2 \cosh\!\left(\frac{V(\phi_h) T}{4}\right)}{V(\phi_h)}
      \right) .
\end{equation}
We now replace the canonical pair $(E,T)$ with $(L,P)$. Starting from
\begin{equation}\label{eq:sympleform}
    \omega = \mathrm{d}T \wedge \mathrm{d}E
    = \mathrm{d}L \wedge \mathrm{d}P ,
\end{equation}
we treat $(E,L)$ as independent variables and regard $T=T(E,L)$ as implicitly defined by \eqref{eq:Lor-rel}. One finds\footnote{Since $\mathrm{d}T = \left(\frac{\partial T}{\partial E}\right)_L \mathrm{d}E
       + \left(\frac{\partial T}{\partial L}\right)_E \mathrm{d}L$, the first term drops out due to the wedge product.}
\begin{equation}
    \mathrm{d}E \wedge \mathrm{d}T
    = \left(\frac{\partial T}{\partial L}\right)_E
      \mathrm{d}E \wedge \mathrm{d}L , \qquad 
    \mathrm{d}P \wedge \mathrm{d}L
    = \left(\frac{\partial P}{\partial E}\right)_L
      \mathrm{d}E \wedge \mathrm{d}L .
\end{equation}
Equating the two expressions yields
\begin{equation}\label{eq:dpdH}
    \left(\frac{\partial P}{\partial E}\right)_L
    =
    \left(\frac{\partial T}{\partial L}\right)_E
    =
    \frac{2}
    {\sqrt{V\!\left(\phi_h(E)\right)^2 - 4 e^{-L}}} ,
\end{equation}
where we used $\phi_h=W^{-1}(16 \pi G_NE) $ to express $V(\phi_h)$ as a function of $E$.
At fixed $L$, the right-hand side depends only on $E$, and integration yields
\begin{equation}\label{eq:pl-H}
    P(H,L)
    =
    2 \int^{H}
    \frac{\mathrm{d}E}
    {\sqrt{V\!\left(\phi_h(E)\right)^2 - 4 e^{-L}}}
    + c(L)
    = \frac{1}{8\pi G_N}
    \int^{W^{-1}(16\pi G_NH)}
    \frac{V(\phi_h)\,\mathrm{d}\phi_h}
    {\sqrt{V(\phi_h)^2 - 4 e^{-L}}}
    + c(L) .
\end{equation}
where in the second step we changed variables from $E$ to $\phi_h$.
The function $c(L)$ is arbitrary; choosing $c(L)\neq 0$ merely corresponds to a further canonical transformation $P \to P +c(L)$. Equation \eqref{eq:pl-H} therefore defines the momentum conjugate to $L$ and, upon inversion, allows one to express the Hamiltonian $H$ in terms of the new variables $(L,P)$ satisfying $\{L,P\}=1$. 

As an example, let us consider JT gravity with dilaton potential
\begin{equation}
V(\phi)= - \Lambda \phi.
\end{equation}
This yields
\begin{equation}
P = - \frac{\sqrt{-4 \tilde\Lambda H-4e^{-L}}}{\tilde\Lambda}, \qquad \tilde\Lambda \equiv \Lambda 8\pi G_N,
\end{equation}
and upon inverting
\begin{equation}
\label{eq:HJT}
H = -\tilde\Lambda P^2 - \frac{1}{\tilde\Lambda}e^{-L}.
\end{equation}
For AdS JT ($\Lambda <0$), this Hamiltonian represents Liouville QM, with positive definite energy eigenvalues \cite{Harlow:2018tqv}. For dS JT ($\Lambda >0$), we have an overall minus sign, and now the energy is purely negative definite.
Since these describe the leading $s$-wave fluctuations around the cold and Nariai regimes respectively, we do indeed match with energy fluctuations only having positive (resp. negative) energy above extremality in both regimes, as in Figure~\ref{fig:sharkfin} above. Note also that setting $\Lambda=0$ is singular in these formulas. We will analyze the flat space JT model shortly.

\subsection{Feature 3: The disk partition function}
One can motivate the following generic expression for the energy spectral density for a generic dilaton gravity \eqref{eq:dila1} \cite{Blommaert:2024whf}:
\begin{equation}
\label{eq:genrho}
\rho_{\text{exact}}(E) = \sum_{\text{preimages }\phi_i(E)} (-)^{\text{sgn}(W'(\phi_i(E))}\frac{1}{8\pi^2}e^{\frac{\phi_i(E)}{4G_N}},
\end{equation}
where we sum over all solutions $\phi_i$ (labeled by $i$) of $E = \frac{1}{16\pi G_N}\int^\phi dx V(x)$ for given $E$.

In case $V(\phi)$ is odd, we can always already combine preimages $\phi_i(E)$ in pairs. This results in the following disk partition function
\begin{equation}
\label{eq:genpfpre}
Z(\beta) = \int_{-\infty}^{+\infty} d\phi_h \, V(\phi_h)\sinh(\frac{\phi_h}{4G_N})e^{-\frac{\beta}{16 \pi G_N} \int^{\phi_h}V(\phi)d\phi},
\end{equation}
valid for AdS JT gravity \cite{Stanford:2017thb}, sinh \cite{Mertens:2020hbs,Fan:2021bwt} and (ungauged) sine dilaton gravity \cite{Blommaert:2024ymv,Blommaert:2024whf,Blommaert:2025avl}.\footnote{Due to different boundary conditions and analytic continuations, this expression is different from the disk amplitudes of the Virasoro minimal string and complex Liouville string \cite{Collier:2023cyw,Collier:2024kmo}.} In the specific case of a monotonic function $W(\phi)$, the answer also simplifies to:
\begin{equation}
\label{eq:genpf}
Z(\beta) = \int_{-\infty}^{+\infty} d\phi_h \, V(\phi_h)\exp(\frac{\phi_h}{4G_N})e^{-\frac{\beta}{16 \pi G_N} \int^{\phi_h}V(\phi)d\phi}.
\end{equation}
In all other cases, we have to stick with \eqref{eq:genrho}
\begin{equation}
Z(\beta) = \int dE \, \rho_{\text{exact}}(E) e^{-\beta E}.
\end{equation}
The expression \eqref{eq:genrho} can be found from the gas-of-defects approach of \cite{Maxfield:2020ale,Witten:2020wvy,Kruthoff:2024gxc} by integrating by parts their final expression for the density of states. It can be motivated from the above general discussion on the canonical quantization of the model. Note that if $W(\phi)$ is a continuous function with $W(\phi\to +\infty) \to +\infty$, $\rho(E)$ is automatically positive everywhere, as necessary for an interpretation as a canonical partition function $Z(\beta)$. 

This is the generic case. In some cases however, the underlying dynamical system has a symmetry in phase space.\footnote{If $V(\phi)$ is a periodic function, there is a discrete shift symmetry in phase space \cite{Blommaert:2024whf}, but the converse statement is not true.} Choosing to gauge that symmetry leads to a different energy spectrum with strong suppressions. This will be relevant for our final iteration of the dilaton gravity model at hand in section \ref{s:ref2}, and we will come back to it.

\section{Starting point: Flat JT gravity}
\label{s:flatJT}
From the dimensional reduction of the Reissner-Nordstr\"om de Sitter black hole, we obtained the two dimensional effective action \eqref{eq:2dgaugefixed}, and are now left with the task of quantizing it properly. We investigate this problem by adopting successive levels of refinement in the treatment of the two dimensional model, with particular emphasis on the spectral density of the system. 
We start with the crudest approximation possible: that of flat space JT gravity with action:
\begin{equation}
\label{eq:flatJT}
S= \frac{1}{16\pi G_N} \int d^2x \sqrt{-g}  \left[\phi R - \Lambda\right].
\end{equation}
The classical metric solution with $\phi$ the radial coordinate is written as:
\begin{equation}
\label{eq:metricflatrind}
     ds^2 = -\Lambda( \phi_h -\phi) {\rm d} t^2 + \frac{{\rm d} \phi^2}{\Lambda (\phi_h -\phi)}.
\end{equation}
For physical transparency, we do \emph{not} yet perform the rescaling of $\alpha$ \eqref{eq:alpha} here, effectively setting $\alpha=1$ in this section. This allows an easier comparison to the $4\mathrm{d}$ black hole system. 

This metric is geometrically flat, has a single horizon at $\phi=\phi_h$, and can be turned into a standard Rindler metric by introducing the proper length $\rho = 2 \frac{\sqrt{\phi-\phi_h}}{\sqrt{-\Lambda}}$:
\begin{equation}
\label{eq:metricflatrind2}
     ds^2 = -\frac{\Lambda^2}{4}\rho^2 {\rm d} t^2 + {\rm d} \rho^2.
\end{equation}
This immediately leads to a temperature and black hole entropy:
\begin{equation}\label{eq:hagedorn}
\beta= -\frac{4\pi}{\Lambda}a, \qquad E 
= -\frac{\Lambda}{16\pi G_N a}\phi_h,
\qquad 
S = \frac{\phi_h}{4G_N} = -\frac{4\pi}{\Lambda} a E,
\end{equation}
where we have reinstated the scale $a$ of the dilaton asymptotics as in \eqref{eq:dilaa}. These $2\mathrm{d}$ equations are to be compared to the (inverse) Hawking temperature of the outer horizon $r_+$ of the near-ultracold $4\mathrm{d}$ black hole \eqref{eq:Tplus}:
\begin{equation}
T_\mathrm{Hawking}(r_+)^{-1} = \frac{6\pi r_0^3}{\lambda^2} = -\frac{8\pi \Phi_0}{\delta_{Q^2}},
\end{equation}
where we used \eqref{eq:Q} for the second expression.
With $\Lambda =  \frac{\delta_{Q^2}}{\Phi_0^2}$ this leads to the identification
\begin{equation}
\label{eq:a}
\boxed{
a = \frac{2}{\Phi_0}},
\end{equation}
as the conversion scale that relates our $2\mathrm{d}$ dilaton gravity model to the original $4\mathrm{d}$ variables, which finishes the identification of the specific dilaton gravity model at hand. With $4\pi G_N = G_4/\Phi_0^2$ ($\alpha=1$ here), we hence get the black hole thermodynamic relations
\begin{equation}
\label{eq:semitdexact}
E = -\frac{1}{16\pi G_N a}\frac{\delta_{Q^2}}{\Phi_0^2}\phi_h, \qquad \beta = -4\pi \frac{\Phi_0^2}{\delta_{Q^2}}a, \qquad S = \frac{\phi_h}{4G_N},
\end{equation}
satisfying $dE = \beta^{-1}dS$.
The on-shell action, or free energy, for flat space JT vanishes:
\begin{equation}
I_{\text{on-shell}} \equiv \beta F = \beta E-S = 0.
\end{equation}
The entropy profile exhibits Hagedorn behavior $S\sim E$ and $\beta$ independent of $E$, which is indeed consistent with the results derived in the CGHS model \cite{Fiola:1994ir,Afshar:2019axx,Stanford:2020qhm,Godet:2020xpk}, related to flat space JT gravity by a Weyl transformation. A thermodynamic system like this has the same temperature for any energy $E$, leading to an infinitely large heat capacity $C= dE/dT$, and one describes this as the thermodynamics being degenerate. Such behavior then usually warrants going to the first subleading correction to obtain a non-trivial $\beta(E)$ thermodynamic relation.\footnote{Hagedorn physics has been well-studied in string theory, see e.g. \cite{Kutasov:2000jp} for a similar motivation to go beyond leading order to break the degeneracy.}

The conjugate momentum to the bulk length $L$ is given by applying \eqref{eq:pl-H}:
    \begin{equation}
        P(H,L)
    =\frac{1}{8\pi G_N}\int^{-16\pi G_N H/\Lambda} \frac{d\phi_h}{\sqrt{\Lambda^2-4e^{-L}}},
\end{equation}
leading to 
\begin{equation}\label{eq:strictflat}
H(L,P) = -2\sqrt{\Lambda^2-4e^{-L}}\, P, \qquad \{L,P\}=1,
\end{equation}
as a function of the two conjugate variables $L$ and $P$. This expression shows a classical bound on the geodesic length $L$ in the auxiliary AdS$_2$ metric $L > -2 \ln \frac{\abs{\Lambda}}{2}$. This classical Hamiltonian $H(L,P)$ is unbounded from above and below in phase space. This is problematic for stability reasons upon quantization.

Indeed, the disk partition function of gravitational fluctuations of flat JT gravity \eqref{eq:flatJT} is of the type given by \eqref{eq:genpf}:
\begin{align}
Z(\beta) &\sim \Lambda\int_{-\infty}^{+\infty} d\phi_h \exp(\frac{\phi_h}{4G_N})e^{\frac{1}{16\pi G_N}\beta \Lambda \phi_h} \\
&\sim 16\pi G_N \int_{-\infty}^{+\infty} dE \exp(-\frac{4\pi E}{\Lambda})e^{-\beta E},
\label{eq:Zflat}
\end{align}
with a spectral density
\begin{equation}
\rho(E) \sim \exp(-\frac{4\pi E}{\Lambda}).
\end{equation}
However, the range of $\phi_h$ and $E$ is $(-\infty,+\infty)$ and this causes the resulting partition function to be badly divergent (depending on sgn$(\Lambda)$ and whether $\beta$ is smaller or larger than $-4\pi/\Lambda$ this can come from either $E\to\pm\infty$ region).\footnote{We note that setting $\Lambda=0$, which corresponds to the zero-temperature saddle, leads to even more singular physics where the spectral density is either zero or infinity for positive or negative energies depending on whether $\Lambda=0^+$ or $\Lambda=0^-$.} This infinite range matches with the role of this model for the ultracold black hole, where the fluctuations in the near-extremal regime can have either sign (unlike the AdS and dS JT cases of relevance for the cold and Nariai regimes discussed above around eq. \eqref{eq:HJT}). Note the classical saddle relation of \eqref{eq:Zflat} leads indeed to the Hagedorn thermodynamics $\beta = - 4\pi / \Lambda$.

We can summarize flat space JT gravity \eqref{eq:flatJT} is a pathological model in the following senses:
\begin{itemize}
\item Its classical thermodynamics is Hagedorn \eqref{eq:hagedorn}, pointing towards the importance of subleading terms in the dilaton potential to relieve this degeneracy.
\item Its energy spectrum with density of states $\rho(E) \sim \exp(-\frac{4\pi E}{\Lambda})$ in \eqref{eq:Zflat} is unbounded from below, and the resulting quantum fluctuations (at disk topology) lead to a divergence.
\item 
We only consider disk topology here, but it is well-known that microphysics is actually contained in higher topologies. For flat JT \eqref{eq:flatJT} however, since $R=0$ in the path integral, one cannot consider surfaces except for the disk and cylinder. This makes flat JT gravity at higher topology an unsatisfactory simple theory.\footnote{This feature however is no longer there beyond the strict JT limit. Indeed, viewing our model as a limit of DSSYK, we can consider taking the flat or high entropic limit directly from a matrix model description of sine dilaton gravity, which is believed to be the ETH-matrix model of \cite{jafferis2022jt}. We will come back to this in the Outlook section \ref{s:concl}.}
\end{itemize}

\section{Refinement 1: Near-flat dilaton gravity}
\label{s:ref1}
The above leading approximation of flat space JT gravity \eqref{eq:flatJT} does not work. In deriving \eqref{eq:2dgaugefixed} however, we carefully tracked the subleading contributions, adding a quadratic term to the dilaton potential with a small prefactor:
\begin{equation}
\label{eq:dilgr}
         S = \frac{1}{16\pi G_N}\int d^2x \sqrt{-g}(\phi R + 2 - \abs{\log q} \phi^2).
\end{equation}
This fundamentally alters the structure of the model and ``regularizes'' some of its problematic features. Here we develop classical and quantum aspects of this dilaton gravity model.

\subsection{Breaking the Hagedorn degeneracy}

The classical solutions of \eqref{eq:dilgr}, in the gauge where the dilaton field $\phi$ is the radial coordinate, but now including the first subleading correction, are given by\footnote{We have here used our rescaled model \eqref{eq:dilgr} for which $\Lambda=-2$. It is very simple to reinstate $\Lambda$ if wanted.}
\begin{equation}
\label{eq:metricflatDSSYK}
     ds^2 = -F(\phi) {\rm d} t^2 + \frac{{\rm d} \phi^2}{F(\phi) }, \qquad F(\phi)  = 2\phi - \frac{\abs{\log{q}}}{3} \phi^3 -16\pi G_N E ,
\end{equation}
where 
\begin{equation}
    E = \frac{1}{16\pi G_N}\int^{\phi_h}d \phi \,V(\phi) 
    = \frac{1}{8\pi G_N}\left(\phi_h - \frac{\abs{\log q}}{6} \phi_h^3\right), 
    \label{eq:energyboundaryvalue}
\end{equation}
is the total energy in the spacetime in terms of the horizon value of the dilaton $\phi_h$. The other thermodynamic quantities are
\begin{equation}\label{eq:thermo2}
\beta= \frac{4\pi}{2 -\abs{\log q}\phi_h^2}
\qquad 
S = \frac{\phi_h}{4G_N}.
\end{equation}
One may then invert the relation \eqref{eq:energyboundaryvalue} in order to find $\beta(E)$ and $S(E)$ in a series expansion in $\abs{\log q}$:
\begin{equation}
\label{eq:hagedorn_correction}
\beta = 2\pi + \pi \abs{\log q}(8\pi G_N)^2E^2 + \mathcal{O}(\abs{\log q}^2),\qquad S = 2\pi E+\frac{\pi(8\pi G_N)^2 \abs{\log q}}{3} E^3 +\mathcal{O}(\abs{\log q}^2).
\end{equation}
These are the inverse Hawking temperature associated with the horizon, and the Bekenstein-Hawking entropy formula for the bulk black hole, satisfying $dE=\beta^{-1}dS$. $S$ and $E$ are normalized such that when the black hole is absent, $\phi_h=0$ and both $S$ and $E$ are zero. These corrections break the Hagedorn degeneracy of flat space JT gravity, and physically resolve one of the pathologies of flat space JT gravity. However, the resulting correction causes $T=\beta^{-1}$ to decrease upon increasing $E$, leading to a negative heat capacity $C=dE/dT<0$, as is usually found for black holes in (approximately) flat spacetime. This is a problem since we are interested in obtaining a thermal partition function for fluctuations, but such an object is ill-defined and divergent for negative heat capacity. This is already a sign that there are still problems with this model.

\subsection{Features of the $2\mathrm{d}$ classical solution}
\label{sub:2d_solution}

In this choice of coordinates, the Ricci scalar throughout spacetime is given by 
\begin{equation}
\label{eq:curvatureFlatDSSYK}
    R = 2\abs{\log{q}} \phi .
\end{equation} 
Taking the $q \to 1 \;(\abs{\log q} \to 0 )$ limit of  \eqref{eq:metricflatDSSYK} or \eqref{eq:curvatureFlatDSSYK} results in a space of solutions describing coordinate patches of flat space with a Rindler horizon at $\phi = 8\pi G_N E$ as described in the previous section. 

\begin{figure}[h]
    \centering
    \includegraphics[width=0.7\linewidth]{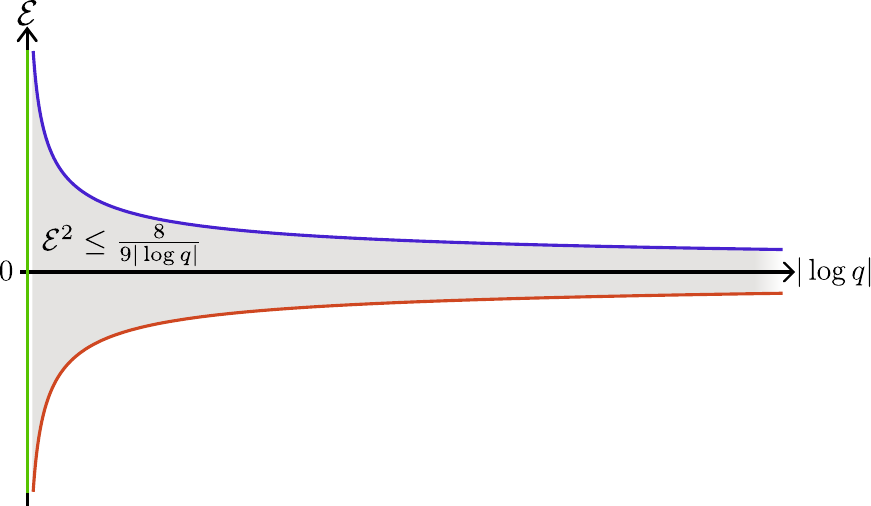}
    \caption{
    Reality of the three horizons constrains the physical solutions of the $2\mathrm{d}$ action to a bounded energy region forming the analog of the sharkfin of figure \ref{fig:sharkfin} in $2\mathrm{d}$. This energy bound grows and diverges as the Rindler space limit is reached at $\abs{\log q} \to 0$.  
    }
    \label{fig:sharkfin2d}
\end{figure}

The metric solution \eqref{eq:metricflatDSSYK} has three horizons 
$\phi_- \equiv \phi_2 < \phi_+  \equiv \phi_1 < \phi_c \equiv \phi_3$ 
located at
\begin{equation}
    \phi_j  = \sqrt{\frac{8}{\abs{\log{q}}}} \cos\left(\frac{1}{3}\left(2 \pi j - \arccos{\left(-24\pi G_N E\sqrt{\frac{\abs{\log{q}}}{8}}\right)}\right)\right), \quad j=1,2,3.
    \label{eq:horizons2d}
\end{equation}
The labeling of the horizons is chosen to reflect that these will be shown to be the remnants of the inner, outer and cosmological horizons of the $4\mathrm{d}$ black hole solution.
Similar to $4\mathrm{d}$, reality of these horizons imposes a bound on the parameters of the system namely, for fixed $\abs{\log q}$ the energy $\mathcal{E}\equiv 8\pi G_N E$ is restricted to
\begin{equation}\label{eq:en_bounds}
\mathrm{Disc}_3 = \frac{4}{3}\abs{\log q}\big(8-9 \mathcal{E}^2 \abs{\log{q}}\big) \geq 0  \implies 
    \frac{9\mathcal{E}^2}{8} \leq 1/\abs{\log{q}},
\end{equation}
leading to a $2$d analog of the sharkfin diagram shown in figure \ref{fig:sharkfin2d}. 
From the blackening factor one readily sees that the energy tunes the coordinate distances between the horizons where saturating these bounds is equivalent to the merging of two of the horizons.\footnote{E.g. at the lower energy bound \begin{equation}
    \phi_j = \sqrt{8} \cos\left(2\pi j/3 \right)/\sqrt{{\abs{\log q}}}
\end{equation} such that $\phi_{1} = \phi_{2}  = -\sqrt{2/\abs{\log q}}$ and similarly for the upper bound $\phi_{1} = \phi_{2}  = \sqrt{2/\abs{\log q}}$.} Note that as $q \to 1$ the bound becomes trivial corresponding to the strict flat JT limit. 

Looking at the expression of the curvature, we see that $\abs{\log q}$ takes the role of a regulator away from the flat limit introducing horizons in the asymptotic region of the spacetime ($\phi_c - \phi_- \sim \abs{\log q}^{-1/2}$). Meanwhile, the energy $\mathcal{E}$ tunes the asymmetry of the horizons, driving the near-horizon region away from flat space toward one of positive (or negative) curvature. This can be seen from the curvature at the outer horizon,
\begin{equation}
    R\vert_{\phi_+} = 2 \abs{\log q} \phi_+ = 2\abs{\log q} \mathcal{E}, \label{eq:R2d}
\end{equation}
which holds for small energies or more generally from the monotonic dependence of $\phi_+$ on $\mathcal{E}$. Furthermore, by plugging in the expression \eqref{eq:horizons2d} for $\phi_-$ and $\phi_c$, the curvature in the entire region between $\phi_-$ and $\phi_c$ is readily seen to be bounded by
\begin{equation*}
    \abs{R} \leq \sqrt{32 \abs{\log q}}. 
\end{equation*}

The statement that the cosmological and inner black hole horizons are located near asymptotic infinity should again be made in terms of the proper distance rather than coordinate distance. One easily finds the proper distance to be
\begin{align}
   L_\mathrm{2d} =& \left(\frac{3}{\abs{\log q}}\right)^{1/2} \int_{\phi_+}^{\phi_c} \frac{dx}{\sqrt{(x - \phi_-)(x-\phi_+)(\phi_c- x)}} \\
   = &\left( \frac{3}{\left|\log q\right|} \right)^{1/2} \left( \frac{1}{\phi_c - \phi_+} \right)^{1 / 2} \int_0^1 \frac{ds}{\sqrt{s\left(1-s  \right)\left( s + \frac{\phi_+ - \phi_-}{\phi_c - \phi_+} \right)  } }.
\label{eq:l2d}
\end{align}
Importantly this calculation reveals that the divergent behavior of the distance between the two horizons is a physical feature of the model \eqref{eq:dilgr}:
\begin{equation}
    L_\mathrm{2d}\: \approx \; \left(\frac{3}{2\abs{\log q}}\right)^{1/4} \sqrt{2}K\left(\frac{\sqrt{2}}{2}\right),
    \label{eq:dist2dleading}
\end{equation}
since $(\phi_c - \phi_+)^{-2} \approx \abs{\log q}/6$ for small $\mathcal{E}$.
We can use these results to quantitatively connect the parameters of the $2\mathrm{d}$ and $4\mathrm{d}$ geometries as follows.
Comparing the length of the throat in the $2\mathrm{d}$ dilaton model \eqref{eq:dist2dleading} with that of the near-ultracold black hole \eqref{eq:dist4dleading} allows for a leading order identification $\lambda$ and $\abs{\log q}$, while comparing the leading order curvature at the outer black hole horizon in both models \eqref{eq:R2d} and \eqref{eq:R4d} yields a relation between the asymmetry of horizons $\delta$ and the ADM energy $\mathcal{E}$ resulting in the leading order dictionary
\begin{align}
    \abs{\log q} = \frac{2 \lambda^2}{3 r_0^6}, \qquad
    \mathcal{E} = -\frac{2\delta r_0^3}{\lambda^2}.
    \label{eq:dict2d4d}
\end{align} 

As we mentioned earlier, when comparing the $4\mathrm{d}$ with the $2\mathrm{d}$ solutions, care has to be taken in identifying the coordinate frames utilized. In particular, in $2\mathrm{d}$ dilaton gravity the classical solutions \eqref{eq:dila} were written in a frame where $\phi$ itself is the radial coordinate. This choice fixes an overall scale of thermodynamic quantities since the asymptotics of the dilaton $\phi$ contains physics on the precise boundary conditions. 
Take the $4\mathrm{d}$ blackening factor \eqref{eq:wfdsdelta} as a starting point and rescale $V(r)$ by $r/r_0$ to match the conventions laid out in \eqref{eq:metric4d}. One then finds that the coordinate transformation
\begin{equation}
\label{eq:variabletrafo}
    r = r_+ + \frac{2\delta}{3} + \frac{3\lambda^2 + \delta^2}{9 r_0^3}\phi , \qquad \tilde t = \frac{9r_0^3}{3 \lambda^2 + \delta^2} t,
\end{equation}
leads to a factorized metric of the form 
\begin{equation}
\label{eq:metricapprox}
    ds^2 \simeq - f_{2\mathrm{d}}(\phi) dt^2 +  f^{-1}_{2\mathrm{d}}(\phi)d\phi^2 + r_0^2 \d\Omega^2 + \mathrm{(subleading)},
\end{equation}
where subleading terms can be computed order by order and are suppressed following \eqref{eq:defultracold} combined with the fact that $\abs{r - r_+}  \leq \lambda \ll r_0 $.\footnote{More precisely the subleading terms are extracted by expanding the blackening factor and outer black hole horizon in terms of $\delta/r_0, \lambda^2/r_0 $ and $(r - r_+)/r_0 = \phi(3\lambda^2 + \delta^2)/9 r_0^4 $.}
 Here the leading contribution to the blackening factor is
\begin{equation}
    f_{2\mathrm{d}}(\phi) = -\frac{4\delta( \delta^2 - 9\lambda^2)r_0^3}{(\delta^2 + 3\lambda^2)^2} + 2\phi - \frac{2(\delta^2 + 3\lambda^2)}{27r_0^6}\phi^3,
    \label{eq:blasckiningfactorexp}
\end{equation}
matching the form of the $2\mathrm{d}$ blackening factor \eqref{eq:metricflatDSSYK} under the identification
\begin{align}
    \abs{\log q} = 
    \frac{2(\delta^2 + 3\lambda^2)}{9 r_0^6}, \qquad 
    \mathcal{E} = 
    2\frac{\delta(\delta^2 - 9\lambda^2) r_0^3}{(\delta^2 + 3\lambda^2)^2}, 
    \label{eq:dict_2ndorder}
\end{align}
where one finds, as required by consistency, the same dictionary as \eqref{eq:dict2d4d} to leading order in the near-ultracold regime $\abs{\delta} \ll \lambda \ll r_0$ and near-horizon regime $\abs{r - r_+} \leq \lambda + \abs{\delta} \ll r_0  $.\footnote{This expansion is valid for all $\abs{\delta} \leq \lambda$, see \ref{ssub:sharkfin2d4d} for some comments.} 
The relevant length scale relating $2\mathrm{d}$ to $4\mathrm{d}$ quantities can be read off from \eqref{eq:variabletrafo}, either as the time rescaling or as the rescaling of the asymptotic behavior of the radial coordinate. It is
\begin{equation}
\label{eq:aaa}
a = \frac{9r_0^3}{3 \lambda^2 + \delta^2} 
= \frac{2}{\alpha \Phi_0},
\end{equation}
and has to be taken into account when converting our dimensionless energies and temperatures from $2\mathrm{d}$ back to dimensionful $4\mathrm{d}$ quantities in terms of the parameters of the near-ultracold black hole itself. This matches our previous determination of this constant by comparing Hawking temperatures in \eqref{eq:alpha} (where we had $\alpha=1$ instead).

\begin{figure*}[h]
    \centering
    \begin{subfigure}[t]{0.5\textwidth}
        \centering
        \includegraphics[width=0.9\linewidth]{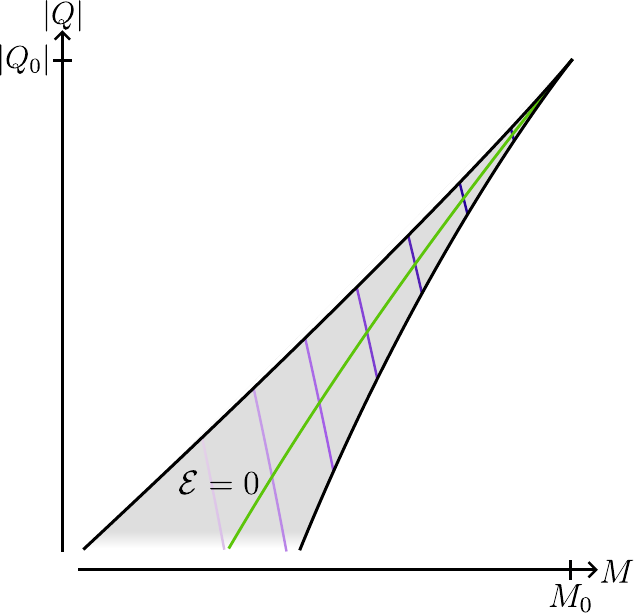}
    \end{subfigure}%
    ~ 
    \begin{subfigure}[t]{0.5\textwidth}
        \centering
        \includegraphics[width=0.9\linewidth]{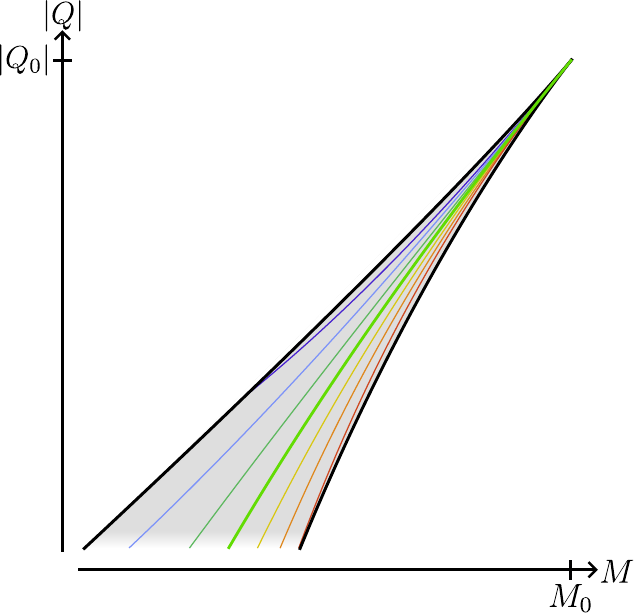}
    \end{subfigure}
    \caption{Plotted are lines at constant $\abs{\log q}$ (left), respectively constant $\mathcal{E}$ (right) for uniformly spaced constant values of $\abs{\log q}$, $\mathcal{E}$ in the $4\mathrm{d}$ sharkfin using \eqref{eq:relation4d2d}. The lines of constant $\abs{\log q}$ are overlaid with the line of constant $\mathcal{E} =0$.  Increasing $\abs{\log q}$ corresponds to larger and larger heating of the ultracold solution i.e. moving further towards the interior of the sharkfin, while varying the energy displaces the solution from the near-ultracold regime (green) towards the cold (blue) or Nariai (red) regime. The green line is the curve at $\delta=0$ on which our reference near-ultracold point is located close to the tip of the diagram.
    } 
    \label{fig:sharkfin4d2d}
\end{figure*}

As a byproduct of this analysis, starting with \eqref{eq:dict_2ndorder} we can also relate the $4\mathrm{d}$ parameters $M$, $Q^2$ to the $2\mathrm{d}$ parameters $\abs{\log q}$, $\mathcal{E}$ including subleading corrections as follows. Using \eqref{eq:outerhorizon}, which we reproduce here for the reader's convenience,
\begin{equation}
    r_+ = -\frac{2 \delta}{3} + r_0\sqrt{1 - \frac{3\lambda^2 + \delta^2}{18 r_0^2}} = -\frac{2\delta}{3} + r_0\sqrt{1 - \frac{\abs{\log q} r_0^4}{4}},
\end{equation}
and the expression derived from the constraints \eqref{eq:coeffrel}
\begin{align}
    Q^2 = \frac{r_+\left((r_+ + \delta)^2 - \lambda^2\right)(3r_+ + 2\delta)}{6r_0^2}, \qquad 
    M = \frac{((2r_+  + \delta)^2 - \lambda^2 )(r_+ + \delta)} {6r_0^2}, 
\end{align}
one finds that $Q^2,M$ naturally arrange in a combination of the expressions \eqref{eq:dict_2ndorder}, which upon invoking the dictionary results in  
\begin{align}
\label{eq:relation4d2d}
    Q^2 &=\frac{1}{2}r_0^2\left(
1-2\abs{\log q}r_0^4
+\frac{7}{16}\abs{\log q}^2r_0^8
-\frac{3}{8}\mathcal{E}\abs{\log q}^2r_0^6\sqrt{4-\abs{\log q}r_0^4}
\right),\\
\label{eq:relation4d2d2}
    M &= \frac{2r_0}{3}\left(
\left(1-\frac{5}{8}\abs{\log q}r_0^4\right)
\sqrt{1-\frac{1}{4}\abs{\log q}r_0^4} + \frac{3}{32}\mathcal{E}\abs{\log q}^2r_0^6
\right).
\end{align}
This rewriting of $M,Q^2$ is exact upon defining the dictionary as \eqref{eq:dict_2ndorder} and thus allows for a bijection of classical solutions in $4\mathrm{d}$ and $2\mathrm{d}$ as we show in \ref{ssub:sharkfinexact}.
Both $M,Q^2$ and the metric then admit a double expansion in the obvious parameters
\begin{align}
   \abs{\log q} r_0^4 = \frac{2(\delta^2 + 3\lambda^2)}{9 r_0^2} \approx \frac{2\lambda^2}{3r_0^2}, \qquad\quad
    \mathcal{E}\abs{\log q}^2 r_0^6 = \frac{8\delta(\delta^2-9\lambda^2)}{81r_0^3} \approx -\frac{8\delta \lambda^2}{9r_0^3}.
\end{align}
Expanding \eqref{eq:relation4d2d} one can invert these relations to find that to leading order in $M,Q^2$
\begin{align}
    \abs{\log q} r_0^4 &\approx 4\left(1-\sqrt{
1+\frac{Q_0^2\delta_{Q^2}}{8r_0^2}
+\frac{3M_0 \delta_M)}{4r_0}
}\right)\approx
 -\delta_M - \frac{\delta_{Q^2}}{8},\\
 \mathcal{E}\abs{\log q}^2 r_0^6 &\approx 8 \left(
1 + \frac{7M_0 \delta_M}{8 r_0} - \frac{3Q_0^2\delta_{Q^2}}{16 r_0^2} 
- \sqrt{1 + \frac{3M_0 \delta_M}{4 r_0}+ \frac{Q_0^2\delta_{Q^2}}{8 r_0^2}} 
\right)\approx \nonumber \\ &\approx  \frac{8\delta_M}{3} -  \delta_{Q^2}.
\end{align}
The near-ultracold point where $\mathcal{E}=0$, is the codimension-1 curve in the sharkfin where $\delta_M = \frac{3}{8}\delta_{Q^2}$, and for which now $\abs{\log q} = -\frac{\delta_{Q^2}}{2r_0^4}$ matching indeed with \eqref{eq:2d4ddictionary}. Using these relations \eqref{eq:relation4d2d}, \eqref{eq:relation4d2d2}, we can visualize the ensembles of constant $\mathcal{E}$ or $\abs{\log q}$ within the $4\mathrm{d}$ sharkfin, as shown in Figure~\ref{fig:sharkfin4d2d}.
Since $\mathcal{E}$ is proportional to the energy deviation in $4\mathrm{d}$ defined by \eqref{eq:semitdexact}, which is in turn conjugate to the temperature of the outer horizon $r_+$, it is tempting to identify in which direction in the sharkfin our energy fluctuations (as described by the dilaton gravity model \eqref{eq:dilgr}) happen; namely along the blue-purple lines of constant $\abs{\log q}$ on the left Figure~\ref{fig:sharkfin4d2d}. We have to be careful however, since quantum fluctuations are off-shell, whereas the sharkfin is on-shell and purely a property of the classical blackening factor. In particular, the charge, as defined by the Coulomb potential \eqref{eq:maxwellsol}, does \emph{not} fluctuate.\footnote{Instead one should, for off-shell configurations, understand $M$ and $Q$ as parameters of the metric disconnected from their thermodynamic definitions. The sharkfin then classifies for which labels a (generally off-shell) configuration contains a naked singularity.}

\subsection{Two-sided phase space description}
We now turn to a Hamiltonian formulation of the regularized flat-space model \eqref{eq:dilgr}. We start from the general expression presented in \eqref{eq:pl-H}, now taking into account the first correction in $\abs{\log q}$ to the dilaton potential: 
\begin{align}
  P( \mathcal{H},\ell)
    &= \frac{1}{8\pi G_N}\int^{ \mathcal{H}+\abs{\log q} \frac{ \mathcal{H}^3}{6}+\cdots} \frac{\d \phi_h \big(2-\abs{\log q} \phi_h^2\big)}{\sqrt{\left(2-\abs{\log q} \phi_h^2\right)^2-4e^{-L}}} \\
    &\approx \frac{1}{8\pi G_N} \int^{\mathcal{H}} \frac{2\d \phi_h}{\sqrt{4-4e^{-L}-2\abs{\log q} \phi_h^2 }} = -\frac{1}{8\pi G_N \abs{\log q}^{1/2}}\arccos \left(\frac{\abs{\log q}^{1/2} \mathcal{H}}{\sqrt{1-e^{-L}}}\right), \nonumber
\end{align}
where we denoted $\mathcal{H}=8 \pi G_{N}H$ and also wrote the inverse of $W(\phi_h)=2\phi_h-\frac{\abs{\log q}\phi_{h}^3}{3}$ perturbatively in $\abs{\log q}$. In the second line, we included the dominant correction to the small $L\approx 0$ regime, which will be the regime of interest in the following. Inverting the resulting relation yields
\begin{equation}
\label{eq:flatH}
\mathcal{H}(L,P)=\frac{\sqrt{1-e^{-L}}}{\abs{\log q}^{1/2}}\cos\left(8\pi G_N\abs{\log q}^{1/2}P\right).
\end{equation}
We notice that expanding~\eqref{eq:flatH} for small $\abs{\log q}$ (after an appropriate shift of the momentum) reproduces the Hamiltonian~\eqref{eq:strictflat} obtained previously in the strict flat-space limit.

We are now interested in a more insightful scaling regime in which the Hamiltonian remains finite while retaining nontrivial dynamics. This is achieved by introducing the rescaled momentum $p=8\pi G_N\abs{\log q}^{1/2}P$. More precisely, if we define the new phase space variables
\begin{equation}\label{eq:rescaled}
L=2\abs{\log q}\ell
\qquad
P=\frac{p}{8\pi G_N\abs{\log q}^{1/2}},
\end{equation}
and expand ~\eqref{eq:flatH} in powers of $\abs{\log q}$, we obtain the finite near-flat Hamiltonian
\begin{equation}\label{eq:Hflat}
\mathcal{H}(\ell,p)=\sqrt{2\ell}\cos(p)+\mathcal{O}(\abs{\log q}).
\end{equation}
The scaling of $L$ in~\eqref{eq:rescaled} is required to obtain a nontrivial and finite Hamiltonian in the limit $\abs{\log q}\to0$. Physically, this small-length regime corresponds to zooming in on the early-time behaviour of the ERB length in the effective geometry. Indeed, specializing Eq.~\eqref{eq:Lor-rel} to the near-flat dilaton potential, rescaling the time coordinate as $T=2\abs{\log q}^{1/2}t$ so as to obtain a nontrivial scaling limit, and matching the leading order terms in $\abs{\log q}$ yields
\begin{equation}\label{eq:l}
\ell=\frac12\left(\phi_h^2+t^2\right),
\end{equation}
which reproduces the well-known quadratic growth of the ERB length at early times.

The induced symplectic form on the phase space \eqref{eq:rescaled} is then
\begin{equation}
\label{eq:flatsimple}
\omega_{\mathrm{flat}}
= dL \wedge dP = \frac{2\abs{\log q}^{1/2}}{8 \pi G_{N}}\d\ell\wedge\d p.
\end{equation}
In other words, the Hamiltonian
\begin{equation}\label{eq:Hflat1}
\mathcal{H}_{\mathrm{flat}}(\ell,p)
=\sqrt{2\ell}\cos p,
\end{equation}
generates the evolution of $\ell$ in \eqref{eq:l} with respect to the rescaled time $t$, as can be verified directly from Hamilton's equations.

However, this Hamiltonian ~\eqref{eq:Hflat1} has an important drawback: it is unbounded both from above and below, so its naive quantization leads to divergent quantum fluctuations and an ill defined quantum theory. We will therefore need a final ingredient to
make sense of the near-flat model at the quantum level, which will be the content of the next Section.

\section{Refinement 2: Gauged dilaton gravity and final proposal}
\label{s:ref2}
The unboundedness of the Hamiltonian \eqref{eq:Hflat1} is problematic both within the two-dimensional effective theory and from the perspective of its four-dimensional uplift. As discussed in subsection~\ref{sub:2d_solution}, sufficiently large positive energies cause the outer and cosmological horizons to coincide, whereas sufficiently large negative energies lead to the merger of the inner and outer horizons. These critical configurations occur at the energy values given in \eqref{eq:en_bounds} and coincide with the energy bounds defining the boundaries of the sharkfin region. Through the identification of these horizons with their four-dimensional counterparts, crossing either of these thresholds is expected to produce a naked singularity in the four-dimensional geometry. As we approach these points, the appropriate low-energy description is expected to be provided by AdS$_2$ or dS$_2$ JT gravity, and our near-flat dilaton gravity model breaks down.

These observations strongly suggest that the quantum theory should contain a mechanism that dynamically suppresses fluctuations whose energies would drive the system beyond these critical values.
Such a mechanism indeed exists. A crucial observation is that  $\mathcal{H}_{\mathrm{flat}}$ \eqref{eq:Hflat1} is invariant under $p \rightarrow p + 2\pi$. 
We propose that the correct quantization of the system is to interpret the invariance of the Hamiltonian $\mathcal{H}_{\mathrm{flat}}$ \eqref{eq:Hflat1} under $p \rightarrow p + 2\pi$ as a gauge redundancy. Since periodically identifying the momentum discretizes its conjugate variable, the conjugate length $\ell$ becomes quantized, $\ell \equiv n$ with $n \in \mathbb{Z}$. In writing this, we have absorbed the prefactors $\frac{8 \pi G_{N}}{2\abs{\log q}^{1/2}}$ of the symplectic form \eqref{eq:flatsimple} into a rescaling of $\hbar$ here to avoid cluttering the equations. We will reinstate them at the end of this section.

Performing the canonical transformation $p\rightarrow p+\pi+\frac{i}{2}\log (2\ell)$, the Hamiltonian \eqref{eq:Hflat1} can be brought to the suggestive non-hermitian form:
\begin{align}
\label{eq:Hflat2}
\mathcal{H}_{\mathrm{flat}}=-\frac12 e^{ip} - \ell e^{-ip}.
\end{align}
Promoting $\ell$ and $p$ to operators satisfying $[\hat{\ell}, \hat{p}] = i$, and keeping the ordering of the variables in \eqref{eq:Hflat2} as written, the Schr\"odinger equation associated with \eqref{eq:Hflat2} for the left and right wavefunctions, defined respectively as $\langle \mathcal{E} \mid \ell \rangle = \psi^{\mathrm{left}}_{\mathcal{E}}(\ell)$ and $\langle \ell \mid \mathcal{E} \rangle = \psi^{\mathrm{right}}_{\mathcal{E}}(\ell)$, is given by:
\begin{align}
\label{eq:schro}
\begin{split}
\frac12 \psi^{\mathrm{left}}_{\mathcal{E}}(\ell-1)+(\ell+1) \psi^{\mathrm{left}}_{\mathcal{E}}(\ell+1)&=-\mathcal{E} \psi^{\mathrm{left}}_{\mathcal{E}}(\ell), \\
\frac12 \psi^{\mathrm{right}}_{\mathcal{E}}(\ell+1)+\ell \psi^{\mathrm{right}}_{\mathcal{E}}(\ell-1)&=-\mathcal{E} \psi^{\mathrm{right}}_{\mathcal{E}}(\ell).
\end{split}
\end{align} 
Due to the discretization of $\ell = n$, these difference equations \eqref{eq:schro} reduce to recurrence relations that naturally truncate to $n\geq 0$ (because $n=0$ is a singular point in the difference equation). In particular, the Hermite polynomials satisfy the three-term recurrence relation
\begin{equation}
\frac12 H_{n+1}(-\mathcal{E}) + n\,H_{n-1}(-\mathcal{E})= -\mathcal{E}\,H_n(-\mathcal{E}), 
\end{equation}
and therefore solve the right eigenvalue equation \eqref{eq:schro} of the regularized flat space model, i.e. $\psi^{\mathrm{right}}_{\mathcal{E}}(\ell)\equiv H_n(-\mathcal{E})$. One can easily show that the left eigenvectors are instead given by $\psi^{\mathrm{left}}_{\mathcal{E}}(\ell)\equiv \frac{H_n(-\mathcal{E})}{2^{n} n!}$. 
The orthogonality relation between left and right eigenvectors is in this case provided by the orthogonality between Hermite polynomials:
\begin{equation}
\label{eq:hermortho}
\int_{-\infty}^{+\infty} \mathrm{d} \mathcal{E}  \ \rho(\mathcal{E}) \psi^{\mathrm{left}}_{\mathcal{E}}(\ell) \psi^{\mathrm{right}}_{\mathcal{E}}(\ell')=\int_{-\infty}^{+\infty} \mathrm{d} \mathcal{E} \, \frac{H_{n}(-\mathcal{E})}{n!  \ 2^{n}}H_{n'}(-\mathcal{E})  e^{-\mathcal{E}^2}=\sqrt{\pi} \delta_{n,n'}.
\end{equation}
One then infers from \eqref{eq:hermortho} the spectral density of the gauged regularized flat space model:
\begin{equation}
\label{eq:gaussian}
\rho(\mathcal{E})= \exp\left(-\mathcal{E}^2\right).
\end{equation}
It is instructive to perform a semiclassical check of this result. 
The thermal partition function is computed by the Euclidean phase space path integral associated to the two-boundary Hamiltonian \eqref{eq:Hflat2}:
\begin{equation}
\label{eq:flat_path}
Z(\beta) = \int \mathcal{D}p \mathcal{D}\ell \exp \left[ \int_{0}^{\beta} \d \tau\left(i p \frac{\d}{\d \tau}\ell+\frac12 e^{ip}+\ell e^{-ip}\right)\right],
\end{equation}
with boundary conditions $\ell(0)=\ell(\beta)=0$. Semi-classically the Euclidean Hamilton equations are
\begin{equation}
i\frac{\d}{\d \tau} \ell = \frac{\partial \mathcal{H}}{\partial p}, \qquad i\frac{\d}{\d \tau} p = -\frac{\partial \mathcal{H}}{\partial \ell},
\end{equation}
or explicitly
\begin{equation}
\frac{\d}{\d \tau}p=-i e^{-ip}, \qquad \frac{\d}{\d \tau}\ell=\ell e^{-ip}-\frac12 e^{ip},
\end{equation}
whose classical solutions are:
\begin{equation}
\label{eq:euclidean}
\ell=-\tau \mathcal{E}-\frac12 \tau^2 \qquad p=-i \log \left(\tau\right),
\end{equation}
unique up to an offset in $\tau$. The on-shell energy associated to \eqref{eq:euclidean} is $\mathcal{H}_{\mathrm{flat}}^{\mathrm{on-shell}}=\mathcal{E}$.

The classical solution for the length $\ell$ in \eqref{eq:euclidean} is constrained to satisfy the periodic boundary conditions $\ell(0)=\ell(\beta)=0$. From this, we immediately obtain
\begin{equation}\label{eq:real_t}
\beta=-2\mathcal{E},
\end{equation}
leading to negative energies $\mathcal{E}<0$ when $\beta >0$. Moreover, we notice the solution for $\ell$ is the Euclidean counterpart of \eqref{eq:l}, obtained by analytically continuing the ERB length to Euclidean time in the usual way via $\tau=\beta/2+it$.

The Euclidean time interval $0\leq \tau \leq \beta=-2\mathcal{E}$ corresponds precisely to the range in which $\ell(\tau)\geq 0$. 
Here this is a property of the classical solution, but it is actually much deeper at the quantum level.
Indeed, imposing that the wavefunction satisfying the Schr\"odinger equation \eqref{eq:schro} has no support on negative lengths $\ell$ automatically selects discretized values for the support of the wavefunction on positive lengths; this is a feature of the Hermite recursion relations. This interplay between length positivity and length discretization is inherited from sine dilaton gravity \cite{Blommaert:2024ymv,Blommaert:2024whf}, and the related q-Hermite recursion relations. It is responsible for the emergence of two notions of temperature in the model, namely $\beta=-2\mathcal{E}$, corresponding to the periodicity of Euclidean time, and $\beta_{\mathrm{fake}}=2\pi$, corresponding to the Hawking temperature of the Rindler horizon.
Given the classical solutions \eqref{eq:euclidean}, one can easily compute the semiclassical approximation to the density of states as the area enclosed by the trajectory in phase space:
\begin{equation}
\rho(\mathcal{E})\simeq \exp\!\left(i\oint p\, \d \ell\right) = \exp \left(-\int_0^{-2\mathcal{E}}\d\tau(\mathcal{E}+\tau)\log\tau\right) =\exp \left(-\mathcal{E}^2\right)\,.
\end{equation}
This expression remarkably coincides precisely with the exact result obtained from the orthogonality properties of the Hermite polynomials \eqref{eq:hermortho}. 

Reinstating the units of $G_N$ and the scale $a$ \eqref{eq:a} to relate $2\mathrm{d}$ to $4\mathrm{d}$ quantities as in \eqref{eq:semitdexact}, one obtains the partition function of asymmetric quantum fluctuations around the near-ultracold black hole:
\begin{align}
\label{eq:propfinal}
Z(\beta) &= \int_{-\infty}^{+\infty} d\phi_h \, e^{-\frac{\phi_h^2}{4\pi G_N}} e^{\beta \frac{1}{16\pi G_N a}\frac{\delta_{Q^2}}{\Phi_0^2}\phi_h} \sim \int_{-\infty}^{+\infty} dE \, e^{-\frac{64 G_4}{\delta^2_{Q^2}} E^2} e^{-\beta E}.
\end{align}
Up to unimportant prefactors, this results in the final density of states
\begin{equation}
\label{eq:dosfinal}
\boxed{\rho(E) = \exp\left(-\frac{64 G_4}{\delta^2_{Q^2}}E^2\right)}.
\end{equation}
The scale of the Gaussian spectral suppression is set by the $4\mathrm{d}$ gravitational constant $G_4$, but it is also sensitive to the (dimensionless) deviation $\delta_{Q^2}$ from the strict ultracold point. We will provide a detailed discussion on the physical lessons of this main result below in section~\ref{s:physconcl}.

\section{From embedding in DSSYK to flat holography}
\label{s:embed}
The above presentation was fully done directly in terms of the dilaton gravity model and how we can make sense of it. Our inspiration on how to do this is from the DSSYK model and its sine dilaton gravity bulk description \cite{Blommaert:2024ymv,Blommaert:2024whf,Blommaert:2025avl}. Here, we reinforce this link, and show that one can view the entire model just as a specific limit of sine dilaton gravity itself \cite{Blommaert:2024whf}. This embedding will be required to make sense of our model in terms of a proposal for flat space holography as we will explain.

\subsection{Embedding in sine dilaton gravity}
The Lorentzian sine dilaton theory is defined by the action:
\begin{equation}
\label{eq:sine_path}
   S = \frac{1}{2 \abs{\log q}}\left[\frac12\int \mathrm{d}^2 x \sqrt{-g}\,\big(\tilde{\phi} R+2\sin\tilde{\phi}\big) + \int \mathrm{d}t \sqrt{-h}\,\big(\tilde{\phi}_b K-\mathrm{i} \,e^{-\mathrm{i} \tilde{\phi}_b/2}\big)\right]\,,
\end{equation}
and has recently emerged as the gravitational bulk dual to the DSSYK model \cite{Blommaert:2024ymv,Blommaert:2024whf,Blommaert:2025avl}. Expanding the action around the maximum of the dilaton potential at $\pi/2$ as $\tilde{\phi} = \pi/2 + \abs{\log q}^{1/2} \phi$ and rescaling the metric as $g_{\mu \nu}=\abs{\log q }^{1/2}g'_{\mu \nu}$, one obtains the near-flat dilaton gravity action \eqref{eq:dilgr}, up to a redefinition of the $2\mathrm{d}$ gravitational coupling constant. 

This embedding is also manifest at the level of classical solutions of sine dilaton gravity. Using \eqref{eq:dila} these are
\begin{align}
        \mathrm{d} s^2&=-\left(-2 \cos\tilde{\phi}+2 \cos\tilde{\phi}_h\right)\mathrm{d} t^2+\frac{1}{-2 \cos\tilde{\phi}+2 \cos\tilde{\phi}_h}\mathrm{d} \tilde{\phi}^2\,. \label{eq:metrrr}
\end{align}
Expanding the radial direction $\tilde{\phi}$ as in the action and accordingly the horizon location $\tilde{\phi}_h$ as 
\begin{equation}
\label{eq:theta_limit}
\tilde{\phi}_h=\pi/2 +\abs{\log q}^{1/2} \mathcal{E}, 
\end{equation}
while performing the same rescaling of the metric with $\abs{\log q}^{1/2}$, one obtains the classical solution \eqref{eq:metricflatDSSYK} of the near-flat dilaton gravity model.

Moving to the quantum level, the canonical quantization of sine dilaton gravity was performed in \cite{Blommaert:2024ymv,Blommaert:2024whf}. As shown in \cite{Blommaert:2024whf}, the near-flat Hamiltonian \eqref{eq:Hflat2} and the associated near-flat path integral~\eqref{eq:flat_path} arise accordingly as the corresponding limit~\eqref{eq:theta_limit} of the sine dilaton gravity Hamiltonian 
\begin{equation}\label{eq:sine_ham}
H_{\mathrm{sine}}=-\cos(P)+\frac12 e^{iP}e^{-L} \, ,
\end{equation}
and the associated q-Liouville path integral \cite{Blommaert:2023wad,Blommaert:2023opb,Bossi:2024ffa}.
As shown in \cite{Blommaert:2024whf}, there are two possible treatments of sine dilaton gravity, depending on whether or not one gauges the emergent $P\rightarrow P+2\pi$ symmetry of the Hamiltonian \eqref{eq:sine_ham}, leading to either a continuum or discretized bulk geometry. 
A useful representation of the partition function, which simultaneously encodes both the ungauged and gauged systems, is
\begin{equation}
\label{eq:gauged_ungauged}
Z_{\mathrm{sine}}(\beta)
=
\int_{-\infty}^{+\infty} d \theta \,
\sin(\theta)\,
\sinh \left(\frac{\pi \theta}{\abs{\log q}}\right)\,
\textcolor{darkblue}{e^{- \frac{\theta^2}{\abs{\log q}}}}\,
e^{\beta \frac{\cos(\theta)}{2 \abs{\log q}}} .
\end{equation}
The blue Gaussian factor distinguishes the two descriptions: its presence reflects the huge reduction of the number of physical states induced by gauging the discrete momentum-shift symmetry, and matches with the DSSYK model partition function \cite{Berkooz:2018jqr}.

The spectral density \eqref{eq:gaussian}, which was our final proposal for the quantization of the near-flat dilaton model, comes directly from the flat space limit \eqref{eq:theta_limit} of gauged sine dilaton gravity (or DSSYK) spectral density \eqref{eq:gauged_ungauged}. Accordingly, the Hermite polynomials arise as the flat-limit \eqref{eq:theta_limit} of the q-Hermite polynomials that diagonalize the transfer matrix of DSSYK.

\subsection{The holographic screen}
Since our resulting model can be interpreted as a model of 2d flat space holography, one obvious question to answer in this framework is that of the location of the holographic screen since different proposals have been put forward in general flat spacetime (see e.g. \cite{Pasterski:2021rjz,Ruzziconi:2026bix} for some recent reviews). Indeed one can directly trace the location of the holographic screen from the limit in sine dilaton gravity, as we show here. This is a particular question where we will need to go beyond the quadratic model \eqref{eq:2dgaugefixed}. Indeed, as the energy $E$ increases and approaches the bounds in \eqref{eq:en_bounds}, or as we geometrically probe locations at large $\phi$, the dimensional reduction performed in \ref{sub:dimred} suggests that progressively more corrections must be included in the effective two-dimensional description. The resulting theory should therefore deviate from the quadratic dilaton model. Nevertheless, although these higher order corrections were neglected in our analysis and are certainly not captured exactly by a sine potential, the sine dilaton embedding remains particularly natural, as it smoothly interpolates between the three known extremal regimes: AdS, flat, and dS JT gravity.  

\begin{figure}
    \centering
    \includegraphics[width=0.5\linewidth]{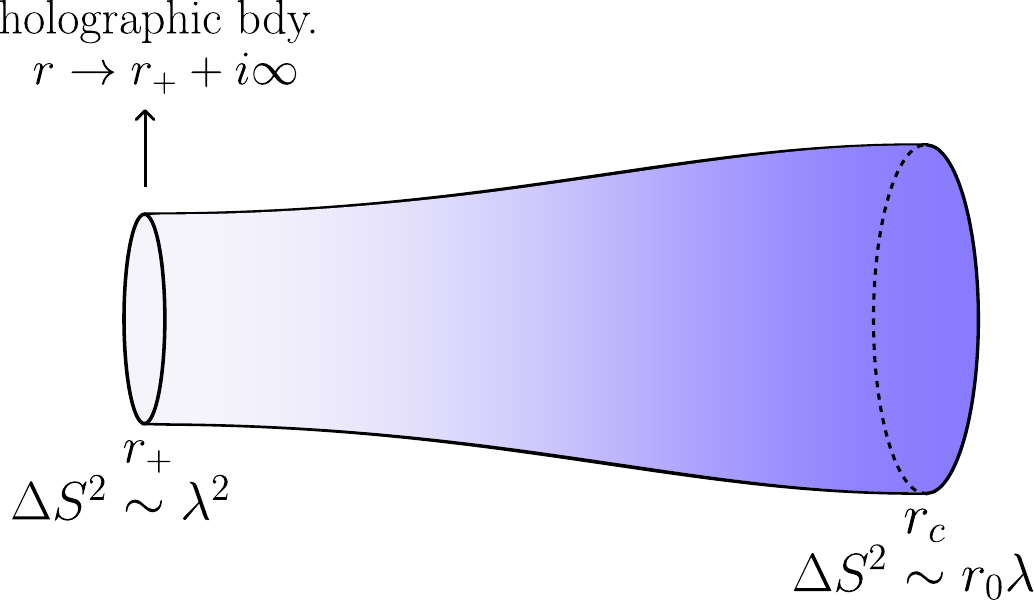}
    \caption{Quantum fluctuations in the near-ultracold $4\mathrm{d}$ black hole are described by a nearly-factorized $4\mathrm{d}$ geometry with a flat near-horizon region and a holographic boundary at $r_+ + i\infty$.}
    \label{fig:geom_hor}
\end{figure}

The conformal rescaling of the metric \eqref{eq:PhiPrime} for sine dilaton gravity is governed by the differential equation
\begin{equation}
\label{eq:odetosolve}
\tilde{\phi}'(\rho)=\frac{2 (\cos\tilde{\phi}_h-\cos\tilde{\phi})}{\rho^2-\rho_{h}^2} \qquad \rho_{h}=\sin\tilde{\phi}_h.
\end{equation}
A solution is given by\footnote{
The general solution of \eqref{eq:odetosolve} is
\begin{equation}
\label{eq:generalsine}
\tilde{\phi}(\rho)=-i \log  \left[\frac{i \sin(\tilde{\phi}_h) \cos \left(\tilde{\phi}_h/2-\kappa\right)+\rho \sin \left(\kappa-\tilde{\phi}_h/2\right)}{i \sin(\tilde{\phi}_h) \cos \left(\tilde{\phi}_h/2+\kappa\right)+\rho \sin \left(\kappa+\tilde{\phi}_h/2\right)}\right],
\end{equation}
and depends on a single integration constant $\kappa$. The case $\kappa=-\tilde{\phi}_h/2$ corresponds to our specific solution \eqref{eq:oursol}. It is basically singled out by the further condition
\begin{equation}
\tilde{\phi}'(\rho)=e^{2\omega}=e^{i \tilde{\phi}(\rho)},
\end{equation}
i.e. that the conformal factor is given by the dilaton itself (or that the effective geometry is described by one of the two Liouville fields in the Liouville formulation), as appropriate for both sine and sinh dilaton gravity. An alternative characterization of this specific choice is that it is the only one that leads to the boundary location at infinite modulus $\abs{\phi}$. Other choices of the integration constant, and the associated location of the holographic screen might be physically interesting for other dilaton gravity models, but we will not explore this here. }
\begin{equation}
\label{eq:oursol}
\phi(\rho)=\frac{\pi}{2}+i \log \left(\rho+i \cos(\phi_h)\right).
\end{equation}
This is the standard boundary condition of sine dilaton gravity, for which the holographic screen is at $\rho\to \infty$ or $\phi \to \pi/2+i \infty$. Since $\abs{\phi}\to\infty$, this screen is very far from where we can trust the series expansion of $V(\phi)=2\sin\phi$ around $\phi=0$, and the nearly flat dilaton gravity model as a good approximation. Defining the nearly flat model as before through the change of variables
\begin{equation}
\label{eq:scalinglimit}
\tilde{\phi}=\pi/2+\sqrt{\abs{\log q}}\phi \qquad \tilde{\phi}_h=\pi/2+\sqrt{\abs{\log q}}\phi_h,
\end{equation}
we deduce that the holographic boundary in terms of $\phi$ is at the locus
\begin{equation}
\phi \to +i\infty,
\end{equation}
or using \eqref{eq:variabletrafo} we can relate this back to the original $r$-coordinate of the $4\mathrm{d}$ RNdS$_4$ black hole:
\begin{equation}
\boxed{
r_{\text{holo. bdy}} = r_+ + i\infty}.
\end{equation}
We conclude that \emph{the holographic boundary of the system of asymmetric fluctuations around the near-ultracold black hole is located at $i\infty$, moving up in the complex plane at the outer black hole horizon $r_+$}, as shown in Figure ~\ref{fig:geom_hor}. As an extra comment, one may also notice that in the original $4\mathrm{d}$ geometry \eqref{eq:dsrnds} the condition $\abs{r} \to \infty$ results in an asymptotic $(t,r)$-geometry proportional to AdS$_2$.

\subsection{Towards flat holography} 
Having identified the location of the holographic screen, we can now insert operators there and compute holographic correlation functions, as a limit of the DSSYK answers. Consider for instance the matrix element $\langle E_f|O_{\Delta, \mathrm{flat}}|E_i\rangle$ of an operator $O_{\Delta, \mathrm{flat}}$ with conformal weight $\Delta$ between energy eigenstates. This matrix element can be found as the corresponding limit \eqref{eq:theta_limit} of the DSSYK matrix element \cite{Berkooz:2018jqr}:
\begin{equation}
\langle \theta_1 | O_\Delta | \theta_2 \rangle
=
\sum_{n=0}^{\infty}
\frac{q^{2n\Delta}}
{(q^2;q^2)_n}\,
H_n(\cos\theta_1|q^2)\,
H_n(\cos\theta_2|q^2)=\frac{(q^{4\Delta };q^2)_\infty}{(q^{2\Delta \pm 2 i\theta_1 \pm 2 i\theta_2};q^2)_\infty},
\end{equation}
or by inserting the bilocal operator $q^{n \Delta}$ into the flat space quantum mechanical model \eqref{eq:Hflat2} as:
\begin{equation}
\begin{split}
\langle E_1 | O_\Delta | E_2 \rangle
&=
\sum_{n=0}^{\infty}
q^{n \Delta}\,
\langle E_1 | n \rangle
\langle n | E_2 \rangle
=
\sum_{n=0}^{\infty}
\frac{q^{n \Delta}}
{2^n n!}\,
H_n(E_1)\,
H_n(E_2) \\
&=\frac{1}{\sqrt{1-q^{2 \Delta}}} \exp \left(\frac{2 q^{\Delta} E_1 E_2-q^{2 \Delta} \left(E_{1}^2+E_{2}^2\right)}{1-q^{2 \Delta}}\right),
\end{split}
\end{equation}
where we inserted the wavefunctions \eqref{eq:schro} and the Mehler's formula for Hermite polynomials was used.

In the bulk, it was shown in \cite{Blommaert:2024ymv} that the dual to $O_\Delta$ is described by a non-minimally coupled scalar field $\varphi$:
\begin{equation}\label{eq:nonminimalc}
-\frac{1}{2}\int d^2x\,\sqrt{-g}\,
\left(
g^{\mu\nu}\partial_\mu\varphi\,\partial_\nu\varphi
+
m^2 e^{-2 i |\log q|\,\tilde{\phi}}\,\varphi^2
\right).
\end{equation}
From the bulk perspective, the non-minimally coupled scalar experiences an effective AdS$_2$ geometry $ds^2_{\text{eff}}$ \eqref{eq:conf_mapping} and behaves near the boundary of this effective spacetime as an ordinary scalar field in AdS$_2$. The effective AdS$_2$ metric is indeed the geometry where canonical quantization is carried out. In the near flat space limit \eqref{eq:theta_limit} and as $\rho \gg 1$, the usual power law asymptotics $\rho^{\pm \Delta}$ of \eqref{eq:nonminimalc} close to the boundary of the effective AdS$_2$ metric become Rindler scattering modes in the real section of the metric $e^{\pm i \Delta \Phi}$ via the mapping \eqref{eq:oursol}. In other words, the modes that are interpreted as non-normalizable and normalizable from the perspective of the effective AdS$_2$ geometry are mapped into incoming and outgoing plane-wave Rindler modes in the original spacetime.

\section{Physical conclusions}
\label{s:physconcl}
\paragraph{Final partition function.}
The resulting final partition function, describing the dominant gravitational fluctuations of the $4\mathrm{d}$ ultracold black hole, is given by
\begin{equation}
\label{eq:Zfinal}
Z(\beta) =\int_{-\infty}^{+\infty} dE e^{-\frac{64 G_4}{\delta^2_{Q^2}} E^2} e^{-\beta E} = \sqrt{\frac{\pi \delta^2_{Q^2}}{64 G_4}}\,e^{\frac{\beta^2 \delta^2_{Q^2}}{256 G_4}},
\end{equation}
with saddle relation $\beta=-128 G_N E/\delta^2_{Q^2}$, and positive heat capacity $C=dE/dT  >0$, and indeed a convergent thermal partition function $Z(\beta)$. A physical partition function with support to arbitrarily negative energies is quite uncommon, especially in a gravitational system. One could expect such a feature in a matrix integral, or in the large $N$ regime of a fermionic system. And indeed, this exact partition function appears in the large $N$ limit of the $O(N)^2 \times O(2)$ fermionic matrix model of \cite{Gaitan:2020zbm}. As already mentioned in the introduction, it also appeared as a toy model of quantum gravity in \cite{Almheiri:2024xtw}, as a direct limit of DSSYK.

Because of the gauging procedure we performed in section~\ref{s:ref2}, and the ensuing vast reduction of states in the Hilbert space of the model, the resulting $2\mathrm{d}$ theory is no longer a theory of $2\mathrm{d}$ black holes. Instead, its fluctuations roughly correspond to observer physics in cosmological spacetimes. One way to appreciate this, is to utilize the embedding of our model within DSSYK, where in \cite{Tietto:2025oxn} a 3d dS observer explanation of this density of states was proposed. A detailed semi-classical $2\mathrm{d}$ bulk understanding of these fluctuations would be interesting, but is not fully clear at the moment. From the $4\mathrm{d}$ perspective, the model is describing quantum gravitational $s$-wave fluctuations of the ultracold black hole with fixed small $\lambda$ (symmetric separation of horizons) and fluctuating $\delta$.
Notice also that the density of states in \eqref{eq:Zfinal} is non-perturbative in $\delta^2_{Q^2}$, and hence we cannot reach the exact ultracold point in perturbation theory after the gauging procedure. 

\paragraph{Small $T$ behavior.}
Unlike the AdS JT partition function, the temperature dependence of the exact disk partition function at low-temperature is highly non-perturbative in $T$. In fact, the series in $T$ (or $G_4$) does not exist since formally
\begin{equation}
\label{eq:weirdanswer}
e^{\frac{\delta^2_{Q^2}}{256 G_4}\frac{1}{T^2}} = \infty +\infty (G_4T^2) + \hdots
\end{equation}
Such divergent series usually appear when a system near a phase transition is expanded in perturbation theory. Here that phase transition point is at $T=0$. 
The free energy and entropy of the exact system \eqref{eq:Zfinal} are given by
\begin{equation}
F = - \frac{\delta^2_{Q^2}}{256 G_4}\frac{1}{T}, \qquad S  = - \frac{\delta^2_{Q^2}}{256  G_4}\frac{1}{T^2}.
\end{equation}
Such behavior is markedly different from the cold and Nariai cases. Indeed, while for the cold and Nariai near-extremal regimes the one-loop correction to the gravitational path integral have been successfully computed, see \cite{Blacker:2025zca} where this computation has been recently presented, for the ultracold near-extremal solution such $4$d result is not known yet. 
The norms of the modes diagonalizing the quadratic operator in the gravitational path integral suffer from both UV and IR divergences and are not normalizable, a puzzle to be solved in the future \cite{uc}. Perhaps this issue is related to the formally infinite expansion coefficients of \eqref{eq:weirdanswer}.

Various results on the one-loop correction are available in the literature, but they don't agree with each other. In \cite{Chen:2025jqm}, a one-loop scaling of $Z\sim T^6$ was proposed for the RNdS$_4$ ultracold black hole, when approaching from the Nariai branch.
Another available result for the one-loop corrections of ultracold black holes was presented in \cite{Arnaudo:2025btb} for the rotating case, where the perturbations are analyzed via the Teukolsky formalism by determining the relevant connection coefficients of the associated Heun equation, leading to $Z \sim T^{3/2}$. Finally, a flat space analogue of JT gravity was considered in \cite{Afshar:2021qvi}. The asymptotic symmetry group in this case is BMS$_2$ and the effective boundary action was called the BMS Schwarzian: it is one-loop exact and controls the low-energy dynamics and thermodynamics of the system in an analogous way to the AdS$_2$ JT gravity case, leading to $Z \sim T^2$. The relation of these works with our results is unclear, especially since our approach based on the gauging procedure of the quadratic dilaton model is so drastically different. 

\paragraph{Bulk discretization.} Our final proposal for the quantization of the near-flat dilaton gravity model relies on gauging the emergent symmetry $p \rightarrow p + 2\pi$ of the gravitational Hamiltonian. As a result, the wormhole length becomes discretized in the bulk. In particular, the lattice spacing is given by
\begin{equation}
\Delta L = 8\pi G_N\abs{\log q}^{1/2} \, \hbar= \sqrt{\frac{8}{\abs{\delta_{Q^2}}}} \frac{G_4}{\Phi_0^2} \, \hbar,
\end{equation}
where, in the second equality, we used the identifications \eqref{eq:2dgaugefixed} and \eqref{eq:2d4ddictionary} relating the two-dimensional and four-dimensional parameters. We emphasize that this lattice is emergent from the quantum dynamics (as indicated by the explicit $\hbar$), and does not exist classically. We note that the lattice spacing increases as $\abs{\delta_{Q^2}} \ll 1$, i.e. as the system approaches the ultracold point. This is consistent with the expectation that the gravitational system becomes increasingly quantum, or equivalently more strongly coupled, as one moves towards the tip of the sharkfin diagram.

\begin{figure}[h]
    \centering
    \includegraphics[width=0.35\linewidth]{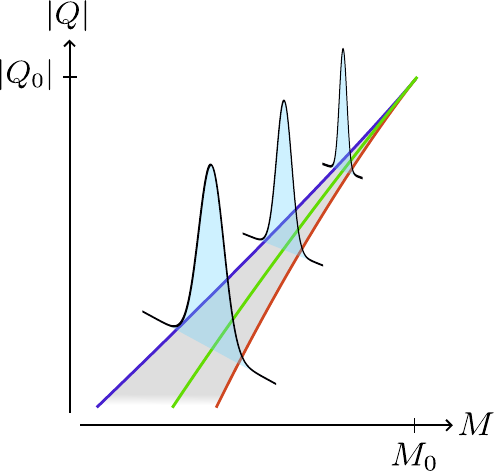}
    \caption{Cartoon of the Gaussian suppression throughout the sharkfin.}
    \label{fig:gaussiansuppression}
\end{figure}

\paragraph{Cosmic censorship.} A remarkable consequence of the gauging procedure performed in section~\ref{s:ref2} is that the Gaussian spectral density derived in \eqref{eq:gaussian} suppresses high-energy fluctuations in both the positive and negative energy directions. From the four-dimensional perspective, this suppression can be interpreted as a manifestation of cosmic censorship operating at the level of quantum fluctuations. Indeed, energy fluctuations exceeding the boundaries of the sharkfin correspond to configurations that give rise to naked singularities in four dimensions. The quantum treatment of the two-dimensional model therefore provides a natural dynamical mechanism that suppresses such pathological configurations. The final spectral density was given by \eqref{eq:dosfinal}:
\begin{equation}
\rho(E) = \exp \left(-\frac{64 G_4}{\delta^2_{Q^2}} E^2\right).
\end{equation}
This Gaussian profile has width $\Delta E= |\delta_{Q^2}|\sqrt{\frac{1}{128 G_4}}$, which shrinks as $|\delta_{Q^2}| \to 0$. Consequently, in the ultracold limit the spectrum becomes increasingly localized around $E=0$. Moving away from the ultracold point and deeper into the sharkfin, the Gaussian broadens, allowing progressively larger energy fluctuations (see Figure \ref{fig:gaussiansuppression} for a sketch of this behavior). This suppression is also manifest at the level of the $2\mathrm{d}$ sharkfin diagram shown in Figure \ref{fig:sharkfin2d}, in terms of $\Delta\mathcal{E} \sim \abs{\log q}^{-1/2}$, increasing as one ventures deeper into the sharkfin away from the strict ultracold point.
This approximate version of \emph{quantum} cosmic censorship seems to be a guiding principle for quantum gravity in any number of spacetime dimensions. Here we encountered it directly in the restricted set of asymmetric horizon fluctuations in $4\mathrm{d}$ quantum gravity. But also in $2\mathrm{d}$ and $3\mathrm{d}$ gravity models, this feature appears. Indeed, at negative cosmological constant, both $2\mathrm{d}$ JT and 3d gravity are governed by an underlying algebraic structure based on SL$^+(2,\mathbb{R})$ \cite{Blommaert:2018iqz} and SL$_q^+(2,\mathbb{R})$ \cite{Mertens:2022ujr}, in terms of a positive (quantum) semigroup \cite{Ponsot:1999uf,Bytsko:2002br,Ip}. This positivity condition ensures only principal series irreps contribute to the Hilbert space of the model. These are closed under tensor product in the (quantum) semigroup, and also ensure no discrete series states (i.e. conical singularities) appear in a complete set of states, which are the lower-dimensional avatars of naked singularities.

\paragraph{Universality.}
We note that the behavior \eqref{eq:weirdanswer}, i.e. $Z(\beta)$ having an essential singularity at $T=0$, is expected generally for a broad class of spectral suppressors that regulate our spectral problem and still support negative energy states.\footnote{We are not aware of such options arising from underlying dynamical considerations, such as gauging a symmetry, but one could be pragmatic and simply postulate a damping factor in the density of states to remove contributions from beyond the sharkfin (naked singularities) as discussed above.} Suppose the partition function to be finite for any finite positive $\beta$ 
\begin{equation}
Z(\beta) = \int_{-\infty}^{+\infty}dE \rho(E) e^{-\beta E}.
\end{equation}
This can be guaranteed by a density of states that acts as a regulator at large energies i.e. $ \forall c \in \R^+ : \abs{\rho(E)} \in o(e^{-c\abs{E}}) $. 
Under this condition the function $Z({\beta})$ is not only finite for positive $\beta$ but uniquely extendable to an analytic function on the complex $\beta$ plane. Assume now that  $0 \neq \rho(E)$ is integrable.  Then by Liouville's theorem $Z(\beta)$ is finite at infinity exactly if it is constant which is excluded since $\rho(E) \neq 0$. Further since $\rho(E)$ is integrable, applying the Riemann-Lebesgue  along $\beta = it $ it follows that $Z(\beta)$ has an essential singularity at $\beta \to \infty$ showing that a large class of regulators indeed produce the expected behavior producing an essential singularity at zero temperature.\footnote{Note that softening the condition that $\rho(E)$ integrable e.g. $\rho(E) \sim \sum_i \delta^{(n_i)}(E_i -E)$ allows for singularities of finite order.}

\label{sec:quantization}
\section{Outlook}
\label{s:concl}

The main motivation behind this paper was the investigation of the quantum properties of the ultracold black hole and its link to $2\mathrm{d}$ dilaton gravity models. In order to find a sensible dilaton gravity model and its quantization, there were quite a few hoops we had to jump through, but in the end we landed on a finite model of flat space holography \eqref{eq:2dgaugefixed}, \eqref{eq:dosfinal}, with physical predictions discussed in the previous section. Here we speculate on where we could take our story next.

\paragraph{Allowing charge fluctuations.} We have throughout this work used a fixed-charge ensemble \eqref{eq:maxwellsol}, where we do not consider independent quantum charge fluctuations. In the case of asymptotically flat near-extremal black holes, charge fluctuations lead to an additional U(1) BF gauge theory in the bulk \cite{Iliesiu:2020qvm}. It would be interesting to work out the precise action describing such fluctuations in our case, possibly as a limit from complex DSSYK and its bulk description \cite{Forste:2025gng,Arundine:2025mcu}. Allowing charge fluctuations would also allow our fluctuations to explore the full $2\mathrm{d}$ sharkfin. We leave this as an important problem for the future. 

\paragraph{A quantum group description of the ultracold black hole.}
In this work we argued that the flat space, or maximal entropy limit of DSSYK yields a consistent quantization of the spherical sector of dynamics governing the ultracold limit of the RNdS$_4$ black hole. One of the prominent ways of describing the quantization of sine dilaton gravity is through the use of quantum groups \cite{Berkooz:2018jqr,Berkooz:2022mfk,Belaey:2025ijg,Schouten:2025tvn,vanderheijdenQuantumSymmetryGeometry2026a}. Specifically it is known that representation theory of a q-deformation of $\mathrm{SL}(2,\mathbb{R})$ not only consistently reproduces partition functions and operator insertions in sine dilaton gravity, but also allows the computation of gravitational wavefunctions in the model. Combining these ideas, it should not come as a surprise that a quantum group description persists in the near-ultracold limit. We claim that the appropriate description is a certain \.In\"on\"u-Wigner contraction \cite{Wigner:1953} of $\mathcal{U}_q(\mathfrak
{sl}_2)$ resulting in a q-deformation of $\mathfrak{iso}(1,1)$. 
The specific q-deformation needs to be chosen such that the discreteness resulting from gauging discussed in section \ref{s:ref2} is preserved in the near-ultracold limit. It would further be interesting to use this limit to better understand the different guises in which quantum groups seem to show up in DSSYK and how this symmetry arises from the isometries of the spacetime under quantization. The model we present here offers a natural setting to study these questions since the flat limit has an emergent isometry in $\mathfrak{iso}(1,1)$, while the quantization leads to discretization in the length variable $\ell$. We view this as an important open question, but leave a detailed investigation to future work. 

\paragraph{Limit of matrix integral for flat space.}
The understanding of higher genus corrections and non-perturbative effects has been one of the most fruitful outcomes in the program of understanding lower dimensional models of quantum gravity. In JT-gravity these effects are captured via the a matrix model completion of the model \cite{Saad:2019lba}. It is known that a similar matrix model completion exists for the DSSYK model  \cite{Jafferis:2022wez,Okuyama:2023kdo,Okuyama:2025xwx} and further that this model can be understood as supplying higher topology contributions to the sine dilaton gravity model. As such it is natural to ask what physics can be extracted from the near-flat limit of this model. Since these corrections have been invaluable for the non-perturbative understanding of quantum information-theoretic quantities like complexity \cite{Iliesiu:2021ari} and entanglement entropy  \cite{Mertens:2025rpa}, it would be interesting to study how the conclusions drawn from these computations generalize beyond the AdS setting and more broadly how higher genus corrections should be viewed in a flat space theory, a hurdle for the flat JT model as mentioned in section \ref{s:flatJT} that gets lifted upon including the corrections and embedding within e.g. DSSYK.

\paragraph{A universal sector of quantum gravity.}
Just as JT gravity is universal in the description of models for near-extremal geometries with two coincident horizons, so is the model described in this work \eqref{eq:2dgaugefixed} for describing the dynamics of black holes near a triple coincident limit of horizons. Assume a gravitational theory to be characterized by a family of classical solutions with at least three horizons such that there exists some region in parameter space of the theory where the subregion of the metric containing these horizons is well approximated by
\begin{equation}
    ds^2 \simeq -f(r)dt^2 + \frac{dr^2}{f(r)} +g(r)dM^2(x^i),
\end{equation}
with the horizons (generically) characterized by simple roots of $f(r)$.\footnote{These assumptions are far from strict and it is expected that with a more in depth analysis this can easily be generalized to more generic setups.}  One then rewrites
\begin{equation}
    f(r) = -(r -r_-)(r - r_+)(r - r_c)h(r),
\end{equation}
with $r_- \leq r_+ \leq r_c$ and three neighboring zeros are selected if $f(r)$ contains more roots. Now by the assumptions of real simple roots $h(r)$ is nonzero between $r = r_-$ and $r = r_c$. Parameterizing the three roots as in \eqref{eq:nearext} and shifting $r \to  x + (r_+ + r_- + r_c)/3 = x + r_+ + 2\delta/3$ yields
\begin{equation}
    f(r) = -\left(\frac{2\delta(\delta^2 - 9\lambda^2)}{27} - \frac{\delta^2 + 3\lambda^2}{3} x + x^3\right)h(r(x)),
\end{equation}
with $x \in \left[-\lambda + \delta/3, \lambda + \delta/3\right]$.
Now assuming $h(r)$ and $g(r)$ to vary slowly between $r_-$ and $r_c$ such that $h(r) \simeq h((r_- + r_+ + r_c)/3)$ and appropriately rescaling $x \to \frac{6}{h(r_+ + 2\delta/3)(\delta^2 + 3\lambda^2)}x$ one immediately reduces the metric to the effective geometry given in \eqref{eq:metricflatDSSYK} with $E \sim \delta(\delta^2 -9 \lambda^2)/(\delta^2 + 3\lambda)^2$ and $\abs{\log q} \sim (\delta^2 + 3\lambda^2)$. Indeed such a geometry always classically develops a long nearly flat throat near $r_+$ in the effective $2\mathrm{d}$ geometry since $R_{2\mathrm{d}} = - f''(r) \simeq 6x h((r_- + r_+ + r_c)/3)$ and $x < \lambda + 3/\delta \to 0$ in the extremal limit. From the perspective of dimensional reduction the expectation is then that such a theory reduces to the $2\mathrm{d}$ effective action \eqref{eq:2dgaugefixed} with the dilaton measuring small fluctuations of the volume of $g(r)dM^2$, under the assumption that only the gravitational subsector of the theory is considered and fluctuations of $dM^2$ in the transverse directions are suppressed.  

\paragraph{$R^2$ gravity.}
A connection we did not at all explore in this work is that to $R^2$ gravity in two dimensions. This connection follows from a further rewriting of the action \eqref{eq:2dgaugefixed} well known in the general dilaton literature for example see \cite{Grumiller:2021cwg,Strobl:1999wv}. Starting from the Euclidean action
\begin{equation}
  I = -\frac{1}{2}\int dx \sqrt{g}(\phi R + 2 - \abs{\log q} \phi^2),
\end{equation}
one notices that the path integral over the dilaton is simply Gaussian (with correct damping from the quadratic term). Thus integrating out the dilaton can be done exactly, resulting in the action
\begin{equation}
    I = -\frac{1}{8 \abs{\log q}}\int dx \sqrt{g} \left( R^2 + 8\abs{\log q}\right).
\end{equation}
This model may be recognized as the torsionless limit of the KV (Katanaev-Volovich) model  \cite{Katanaev:1986wk,Katanaev:1986qm}. As a simple dilaton model it has been studied in different settings both in the classical and quantum regime see e.g. \cite{Kazakov:1996zm,Knizhnik:1988ak,Kawai:1993np}. It would be interesting to investigate what lessons can be learned from combining this rewriting with the quantization proposed in our work.

\section*{Acknowledgments}
We thank Matthew Blacker, Andreas Blommaert, Alba Grassi, Luca Griguolo, Cristoforo Iossa, Lorenzo Russo, Watse Sybesma, Alex Tarana and Chiara Toldo for discussions. FM (doctoral fellowship 11P9Z24N) and TT (doctoral fellowship 11A2925N) are supported by a Research Foundation - Flanders (FWO) doctoral fellowship. TM and JP acknowledge financial support from the European Research Council (grant BHHQG-101040024). Funded by the European Union. Views and opinions expressed are however those of the author(s) only and do not necessarily reflect those of the European Union or the European Research Council. Neither the European Union nor the granting authority can be held responsible for them.

\appendix

\section{Beyond the center of the sharkfin}
\label{ssub:sharkfin2d4d}
In the paper we focused on describing the physics close in what we define to be the near-ultracold limit of the RNdS$_4$ black hole. We defined this limit in \eqref{eq:defultracold}. 
While the physical conclusions of the quantized model we present require this triple hierarchy of scales $\abs{\delta} \ll \lambda \ll r_0$ it turns out that a lot of the classical analysis holds on more general grounds. In this appendix we comment on the validity of results beyond this strict limit while leaving a detailed analysis beyond this regime to future work. 

\subsection{The cold and Nariai regimes}
One straightforward softening of regimes is achieved by only imposing 
 \begin{equation}
    \abs{\delta} \leq \l \ll r_0.
\end{equation}
This limit leads to a model describing the full tip of the sharkfin i.e. geometries with small curvatures $R_{2\mathrm{d}} \in \mathcal{O}(\lambda/r_0^3)$ in the region between the inner and cosmological horizons, where however the separation of curvature scaled between the different horizons breaks down generically. Of course this is expected since this limit should still include Nariai and cold black holes. 
Indeed this is directly visible in the dimensionally reduced geometry \eqref{eq:metricapprox}. For example it almost follows trivially that expanding the blackening factor for $\delta = \pm(\lambda - \epsilon)$, i.e. where the outer black hole horizon merges with either the cosmological or inner black hole horizon, leads to a near (A)dS$_2$ geometry upon imposing $\epsilon \ll \lambda$. Combining this with a similar argument for the dimensionally reduced action \eqref{eq:newdilaction} one thus finds that softening the limit $\abs{\delta} \ll \lambda$ allows the classical model \eqref{eq:2daction} to interpolate between physics at all three extremal limits of the RNdS$_4$ black hole (see section \ref{sec:RNdS} for the various definitions). This poses the obvious question if there exists an extension of the quantization performed in this work to the full tip of the sharkfin. We leave this for future work, however notice that such a quantization should incorporate both features of this model, such as discreteness of wormhole lengths, and features of the other extremal limits, namely an explicit boundedness of the Hamiltonian. 
\subsection{The sharkfin of the effective geometry}
\label{ssub:sharkfinexact}
In the previous subsection we argued that most of the classical analysis can be extended beyond near-ultracold limit as long as $\lambda \ll r_0$. It turns out, however, that some features persist even more generically, most prominently the mapping of physically allowed configurations induced by \eqref{eq:dict_2ndorder}. 
While the dimensionally reduced metric \eqref{eq:metricflatDSSYK} is only an approximation of the full geometry \eqref{eq:metric4d}, this is not the case for the defining relations of the sharkfin \eqref{eq:4ddisc}. Indeed rewriting \eqref{eq:4ddisc} in terms of $E,\abs{\log q}$ using \eqref{eq:relation4d2d} we find
\begin{align}
    \mathrm{Disc}_4 =& \abs{\log q}^3(8 - 9E^2\abs{\log q}) P(\abs{\log q},E) = \frac{3}{4}\mathrm{Disc}_3 \abs{\log q}^2P(\abs{\log q},E),
\end{align}
where $P(E,\abs{\log q})$ is strictly positive if $\abs{\log q} < \frac{4}{9 r_0^4}$.\footnote{For completeness $ P(\abs{\log q},E) =2-\frac{15\abs{\log q}}{8}+\frac{297\abs{\log q}^2}{512}-\frac{3E\sqrt{4-\abs{\log q}}\abs{\log q}^2}{128}-\frac{121\abs{\log q}^3}{2048}+\frac{33E\sqrt{4-\abs{\log q}}\abs{\log q}^3}{4096}+\frac{9E^2\abs{\log q}^4}{32768}$.} Note that this bound on $\abs{\log q}$ is separate from the condition that $\abs{\log q}$ is small required to be close to the tip of the sharkfin and indeed can be shown to be related to assuring $M,Q^2$ to be real and positive. In this region one thus finds the positivity of $\mathrm{Disc}_3$ to be both a necessary condition and sufficient condition for a solution to lie in the full $4\mathrm{d}$ moduli space. 
This bijection of classically allowed points in the solution spaces of the two theories is expected for the following reason: 

Firstly note that the sign of a discriminant is generally preserved under projective (and as such affine) transformations like the one defined in \eqref{eq:variabletrafo}. Further for any quartic polynomial the number of real roots is real. As such that guaranteeing the reality of any three of the roots directly implies reality of the last.

Now to ensure reality of all horizons consider $V(r)$ in the form \eqref{eq:wfds1} applying the coordinate transformation \eqref{eq:variabletrafo} to find\footnote{The factor $d\phi^2/dr^2$ is included since $V(r)$ transforms like a metric component not a polynomial, however since the discriminate is homogeneous under rescaling this does not effect the argument.}
\begin{align}
    V(r)\frac{d\phi^2}{dr^2} &= \left(-\frac{4\delta(\delta^2 - 9\lambda^2)r_0^3}{(\delta^2 + 3\lambda^2)^2} + 2\phi - \frac{2(\delta^2 + 3\lambda^2) \phi^3}{r_0^6}\right)\left(\frac{4}{r_0} \frac{(r(\phi) -r_n)}{r(\phi)^2}\right) =\\
    &=f_\mathrm{2d}(\phi)\left( \frac{4}{r_0} \frac{(r(\phi) -r_n)}{r(\phi)^2}\right).
\end{align}
where we've left the term $(r(\phi)- r_n)/r(\phi)^2$ implicit in $\phi$. Thus the exact mapping of sharkfin boundaries follows from the fact that since all subleading corrections to \eqref{eq:metricapprox} are due to the expansion of $\left(r(\phi)-r_n\right)/r(\phi)^2$ such that the other three roots are exactly preserved. Reality of the three roots $r_-,r_+,r_c$ is then equivalent to reality of the roots of $f_\mathrm{2d}(\phi)$ together with the fact that reality of $r_n$ is guaranteed by reality of the other roots.\footnote{In more technical language this uses the behavior of the discriminant under factorization \cite{GelfandKapranovZelevinsky1994} $\mathrm{Disc}(g) = \mathrm{Disc}(f)\mathrm{Res}(f,h)^2\mathrm{Disc}(h)$, where in our case $V(r)r^2 = g(x) = f(x)h(x) = f_\mathrm{2d}(r)(r -r_n)\alpha$ for appropriate $\alpha$ and the fact approximation only affects $\mathrm{Res}(f_\mathrm{2d},h)^2\mathrm{Disc}((r -r_n)\alpha)$ but not $\mathrm{Disc(f_\mathrm{2d})}$.} Note that from this perspective the shift in the coordinate transformation \eqref{eq:variabletrafo} is simply the centroid of the three physical roots guaranteeing that $f_\mathrm{2d}(\phi)$ has no quadratic term required by the gauge choice of \eqref{eq:metricflatDSSYK} while the scaling normalizes the linear term of $f_\mathrm{2d}(\phi)$.

As such we find that the mapping derived in this work gives a bijection of a large subregion of physically allowed solutions far beyond the limits imposed in the near-ultracold regime. Of course this bijection does not extend to the action of these configurations or especially the quantization discussed in the main text, however still hints that certain features observed in the ultra cold limit may, at least qualitatively, hold far beyond the regime considered.

\bibliographystyle{ourbst}
\bibliography{Refs}
\end{document}